\documentclass[conference]{IEEEtran}
\IEEEoverridecommandlockouts
\def\BibTeX{{\rm B\kern-.05em{\sc i\kern-.025em b}\kern-.08em
    T\kern-.1667em\lower.7ex\hbox{E}\kern-.125emX}}

\usepackage[T1]{fontenc}
\usepackage[utf8]{inputenc}
\usepackage{amsmath,amsfonts,amssymb}

\usepackage{thm-restate}
\usepackage{amsthm}
\usepackage{bm}
\usepackage{extarrows}
\usepackage{stmaryrd}
\usepackage{proof}
\usepackage{bussproofs}
\usepackage[ruled,vlined,linesnumbered]{algorithm2e}
\usepackage{algpseudocode}
\usepackage[usenames, table, svgnames, dvipsnames]{xcolor}

\usepackage{makecell}

\usepackage{tabularx}
\usepackage{graphics}

\usepackage[usenames, table, svgnames, dvipsnames]{xcolor}
\usepackage{url}
\usepackage[hyperfootnotes=true,colorlinks,linkcolor=RedViolet,anchorcolor=black,citecolor=blue,urlcolor=black,breaklinks=true]{hyperref} 

\usepackage{orcidlink}
\usepackage{tikz}
\usetikzlibrary{arrows,shapes,snakes,automata,backgrounds,petri,positioning}
\usepackage{diagbox}
\newcommand{\oomit}[1]{}

\newcommand{\tr}[1]{\langle #1 \rangle} 
\newcommand{\semantic}[1]{\left\llbracket #1 \right\rrbracket} 
\newcommand{\cbrackets}[1]{\left\{ #1 \right\}} 
\newcommand{\pbrackets}[1]{\left( #1 \right)} 
\newcommand{\abrackets}[1]{\left\langle #1 \right\rangle} 
\newcommand\hole{\heartsuit}

\usepackage{todonotes}
\newcommand{\aj}[1]{\todo[inline,color=green!10]{\textbf{AJ:} #1}}

\newcommand{\znj}[1]{\todo[inline,color=orange!10]{\textbf{ZNJ:} #1}}

\usepackage{orcidlink}

\newcommand{\myparagraph}[1]{\smallskip\noindent{\bf #1.}}
\allowdisplaybreaks

\spnewtheorem{mytheorem}{Theorem}{\bfseries}{\rmfamily} 
\spnewtheorem{myproblem}{Problem}{\bfseries}{\rmfamily} 
\spnewtheorem{mylemma}{Lemma}{\bfseries}{\rmfamily} 
\spnewtheorem{myexample}{Example}{\bfseries}{\rmfamily}
\spnewtheorem{myassumption}{Assumption}{\bfseries}{\rmfamily}
\spnewtheorem{mydefinition}{Definition}{\bfseries}{\rmfamily}
\spnewtheorem{myremark}{Remark}{\bfseries}{\rmfamily}

\begin{document}

\title{On Synthesis of Timed Regular Expressions}

\author{\IEEEauthorblockN{Ziran Wang\orcidlink{0009-0004-5555-5932}}
\IEEEauthorblockA{
\textit{Key Lab. of System Software (CAS)} \\
\textit{Institute of Software, Chinese Academy of Sciences,} \\
\textit{\& University of Chinese Academy of Sciences} \\
Beijing, China \\
wangzr@ios.ac.cn}
\and
\IEEEauthorblockN{Jie An\textsuperscript{*}\orcidlink{0000-0001-9260-9697}}
\IEEEauthorblockA{\textit{National Key Lab. of Space Integrated Information System} \\
\textit{Institute of Software, Chinese Academy of Sciences,}\\
\textit{\& University of Chinese Academy of Sciences} \\
Beijing, China \\
anjie@iscas.ac.cn}
\and
\IEEEauthorblockN{Naijun Zhan\textsuperscript{*}\orcidlink{0000-0003-3298-3817}}
\IEEEauthorblockA{\textit{School of Computer Science \& MoE Key Lab.} \\ \textit{of High Confidence Software Technologies} \\
\textit{Peking University}\\
Beijing, China \\
njzhan@pku.edu.cn}
\and
\IEEEauthorblockN{Miaomiao Zhang\orcidlink{0000-0001-9179-0893}}
\IEEEauthorblockA{\textit{School of Computer Science and Technology} \\
\textit{Tongji University}\\
Shanghai, China \\
miaomiao@tongji.edu.cn}
\and
\IEEEauthorblockN{Zhenya Zhang\orcidlink{0000-0002-3854-9846}}
\IEEEauthorblockA{
\textit{Kyushu University}\\
Fukuoka, Japan. \\
zhang@ait.kyushu-u.ac.jp}

\thanks{*~Corresponding authors.}
}


\maketitle

\begin{abstract}
Timed regular expressions serve as a formalism for specifying real-time behaviors of Cyber-Physical Systems. In this paper, we consider the synthesis of timed regular expressions, focusing on generating a timed regular expression consistent with a given set of system behaviors including positive and negative examples, i.e., accepting all positive examples and rejecting all negative examples. We first prove the decidability of the synthesis problem through an exploration of simple timed regular expressions. Subsequently, we propose our method of generating a consistent timed regular expression with minimal length, which unfolds in two steps. The first step is to enumerate and prune candidate parametric timed regular expressions. In the second step, we encode the requirement that a candidate generated by the first step is consistent with the given set into a Satisfiability Modulo Theories (SMT) formula, which is consequently solved to determine a solution to parametric time constraints. Finally, we evaluate our approach on benchmarks, including randomly generated behaviors from target timed models and a case study.
\end{abstract}

\begin{IEEEkeywords}
Timed regular expressions, Synthesis, Passive learning, Real-time behaviors, Cyber-physical systems
\end{IEEEkeywords}

\section{Introduction} \label{sc:introduction}

In their seminal work~\cite{AsarinCM02}, Asarin et al. introduced timed regular expressions (TRE) as a formalism for specifying real-time behaviours of timed systems, and demonstrated the equivalent expressiveness of generalized extended TRE with timed automata (TA)~\cite{Alur94} proposed by Alur and Dill. 
Their relationship reflects the classic correspondence between finite-state automata (FSA) and regular expressions (RE), where both formalisms describe the same class of languages but offer complementary advantages. TAs excel at modeling stateful transitions with timing constraints, whereas TREs provide a concise, compositional syntax for specifying timed patterns. This duality not only underscores the theoretical expressiveness of TREs but also motivates their practical utility
in learning and synthesis, such as REs have been widely adopted for pattern inference in untimed settings. 

As known, TA is one of the most popular and widely used formal models for real-time systems. For scheduling analysis, as discussed in \cite{DBLP:books/sp/22/TangG022} \cite{Ekberg2020}, most of the task models are subclasses of TA (see Figure 4 in \cite{DBLP:books/sp/22/TangG022}). Specifically, a Di-gragh~\cite{StiggeEGY11} characterized task can be modeled by a labeled time automaton (i.e., a task automaton in \cite{DBLP:books/sp/22/TangG022}), by introducing a clock recording the separation time of two successive jobs. 

TREs have also been found numerous applications in modeling and monitoring of real-time systems and cyber-physical systems, similar to how regular expressions play a key role in processing text and system logs. For instance, the timed pattern matching problem~\cite{UlusFAM14,WagaAH16,WagaH18,WagaAH23} involves identifying all sub-segments of a signal or system trace that match a given TRE which serves as a timed pattern for real-time specifications. There are also several applications to mining TRE-formed specifications from traces of the QNX Neutrino RTOS and an automobile CAN bus~\cite{SchmidtNF17,NarayanCJF18}. 

However, these works, both for scheduling analysis and pattern matching, rely on the assumption that the models are known beforehand. Therefore, for analyzing black-box systems or gray-box systems, where the models are unavailable, a key challenge is to learn a formal model from observable real-time system behaviors. For instance, consider a gray-box task system where we know the type of task model, but the parameters (e.g., the deadline and WCET of jobs) are unknown, and we want to confirm the parameters. The tasks are running on a (or multiple) processor(s) with a certain scheduling strategy, and we can only access the traces of processor behaviors: receiving a job, preemption, completion of a job, and when these behaviors are taken. As shown in \cite{DBLP:books/sp/22/TangG022}, a task system can be modeled by a TA. To determine the unknown parameters, we can infer or synthesize a formal model (e.g., a TA or a TRE) from the observed behaviors. Section~\ref{sc:motivating} provides an illustrative example for the Earliest Deadline First (EDF) scheduling in a gray-box setting.

In recent decades, passive learning involving synthesizing formal models from system behaviors has garnered significant interest. Prior research has proposed methods to automatically generate regular expressions from examples~\cite{Angluin78,Gulwani11,LeeSO16,PanHXD19}, natural languages~\cite{KushmanB13,LocascioNDKB16,ZhongGYPXLLZ18} and multiple modals~\cite{Chen0YDD20}. Learning finite-state automata from examples also attracted attention since 1970s~\cite{Gold78,RivestS93,ParekhH01}. Alternatively, in~\cite{Angluin87}, Angluin proposed an active learning algorithm named $L^*$ within the Minimally Adequate Teacher (MAT) framework, where a learner learns a regular language from a teacher by making membership and equivalence queries.
For timed models, both passive and active learning efforts have primarily concentrated on learning various subclasses of timed automata, e.g., event-recording automata~\cite{Grinchtein10,HenryJM20}, real-time automata~\cite{VerwerWW12,SchmidtGHK13,CornanguerLRT22,AnWZZZ21,AnZZZ21}, timed automata with one-clock~\cite{VerwerWW11,AnCZZZ20,XuAZ22}, Mealy machine with one timer~\cite{DierlHKKLLM23,VaandragerB021}, deterministic timed automata~\cite{TapplerALL19,AichernigPT20,TapplerAL22,Waga23,TengZA24}, etc. Recently, an inference algorithm for timed partial orders (TPO) was presented in~\cite{WatanabeFHLP0023}. TPO can also be viewed as a subset of timed automata. 

However, the subclasses of timed automata mentioned above are deterministic. Therefore, existing learning or synthesizing algorithms for these models cannot be directly applied to the synthesis of TRE involving nondeterministic behaviors. The reduction from TA to TRE (i.e., without renaming function) may not always be possible.
Moreover, for deterministic cases,  synthesizing TA is irrelevant to 
 minimizing TREs. The latter itself is a difficult problem, even for traditional regular expressions,  it is still  PSPACE-complete~\cite{MeyerS72}.

\myparagraph{Contributions} In this paper, we consider the synthesis of TRE from system behaviors, highlighting the following several distinctions from the extensively studied synthesis problem of traditional regular expressions (RE). 
\begin{itemize}
    \item We establish the decidability of synthesizing a TRE consistent with a given set of timed words, which includes positive and negative examples, by exploring a notion of simple timed regular expressions and their falsifiability. 
    \item Subsequently, we propose a method for a particular synthesis problem known as minimal synthesis, aiming to generate a consistent TRE with minimal length. The synthesis process begins by generating a template and then determining the timing constraints. Specifically, in the first step, we generate an untimed version of TRE, referred to as parametric TRE, serving as the template. To achieve a TRE with minimal length, we systematically enumerate such templates starting from a length k.
    \item Additionally, we introduce two strategies for pruning unnecessary candidates to expedite the enumeration. In the second step, we encode the requirement that a candidate parametric TRE remains consistent with the given set into a formula in Satisfiability Modulo Theories (SMT)~\cite{BarrettSST21}, which is then solved to determine a solution to the parametric timing constraints within the parametric TRE. If no solution is found, the process jumps to the first step. We provide the proof of termination and correctness of the synthesis process at the end.
    \item Finally, we implement and evaluate our synthesis process on benchmarks, including randomly generated behaviors of target timed models and a case study.
\end{itemize}

\myparagraph{Outline} Section~\ref{sc:motivating} provides a motivating example, and then  Section~\ref{sc:preliminary} recalls some basic notions of  timed regular expressions. The TRE synthesis problem ia formalized in Section~\ref{sc:problems}, and then the decidability is shown in Section~\ref{sc:decidability}. The two-step synthesis process is presented in Section~\ref{sc:syn_pro} and evaluated in Section~\ref{sc:experiments}. Finally, Section~\ref{sc:conclusion} concludes the paper. 

The omitted proofs of the lemmas and theorems are available in the full version~\cite{wang2025arXiv}.


\section{Illustrative Example} \label{sc:motivating}

To illustrate a potential motivation for synthesizing formal models from the execution traces of real-time systems, we present the following application scenario:
Consider a task model where the minimum interarrival time and deadline are partly unknown and are determined by external factors. However, the execution time is controllable and can be influenced by the resources allocated to the task. For instance, if the tasks are requests from a website, the execution times will depend on the server resources allocated by the service provider. Before commencing the formal operation of the system, it is crucial to determine the execution time required to ensure the schedulability of the system. To achieve this, trial runs are conducted to analyze the parameters of the minimum interarrival time and deadline from the execution traces. 

\begin{figure}[!t]
    \centering
    \resizebox{1.0\linewidth}{!}{
    \begin{tikzpicture}[->, >=stealth', shorten >=1pt, auto, node distance=5cm, semithick,scale=0.6, every node/.style={scale=0.8}, initial where=above]
    \node[initial,accepting, state] (0) at (-9,0) {\large $q_0$}; 
    \node[state] (1) at (-3,3) {\large $q_1$};
    \node[state] (2) at (-3,-3) {\large $q_2$};
    \node[state] (3) at (3,0) {\large $q_3$};
    \node[state] (4) at (3,6) {\large $q_4$};
    \node[state] (5) at (3,-6) {\large $q_5$};
    \node[state] (6) at (9,0) {\large $q_6$};
    \node[state] (7) at (9,-6) {\large $q_7$};
    \node[state] (8) at (9,6) {\large $q_8$};
    \node[state] (9) at (-9,6) {\large $q_9$};
    \node[rectangle, minimum width=3cm, minimum height=2cm] (box) at (-2,0) {\makecell{\large $a_1,x_1\geq p_1\wedge$ \\$ y_2 \geq D_2 -D_1,\{x_1,y_1\}$}};
    \node[rectangle, minimum width=3cm, minimum height=2cm] (box) at (11,4) {\makecell{\large $a_1,x_1\geq p_1\wedge$\\$y_2<D_1-D_2$,\\$\{x_1,y_1\}$} };
    \node[rectangle, minimum width=3cm, minimum height=2cm] (box) at (-7.5,5) {\makecell{\large $c,\top,\{\}$}};
    \node[rectangle, minimum width=3cm, minimum height=2cm] (box) at (11,-1.5) {\makecell{\large $a_1,x_1\geq p_1\wedge$ \\$ y_2 \geq D_2 -D_1,$ \\$\{x_1,y_1\}$}};
    \node[rectangle, minimum width=3cm, minimum height=2cm] (box) at (11,-4.5) {\makecell{\large $a_2,x_2\geq p_2$,\\$\{x_2,y_2\}$}};
    
    \draw [thick, rounded corners=0.4cm] (-6.5,-8) rectangle (13,9);
    \draw [] (-6.5,6) -- (9);
    \path 
    (0) edge [in = -150, out = -60] node [below left] {\large $a_2,x_2\geq p_2,\{x_2,y_2\}$} (2)
    (2) edge[in = -30, out = 150] node [below]{\large $b_2,\top,\{\}$} (0)
    (0) edge [in = 150, out = 60] node [above left] {\large $a_1,x_1\geq p_1,\{x_1,y_1\}$} (1)
    (1) edge[in = 30, out = -150] node [above]{\large $b_1,\top,\{\}$} (0)
    (2) edge[in = -90, out = 00] node [below ]{\makecell{\large $a_1,x_1\geq p_1\wedge$ \\$ y_2 < D_2 -D_1,\{x_1,y_1\}$}} (3)
    (2) edge[in = -90, out = 60] node [below left]{} (4)
    (2) edge[in = 180, out = -90] node [below left]{\large $a_2,x_2\geq p_2,\{x_2,y_2\}$} (5)
    (1) edge[in = 150, out = 0] node [left]{\makecell{\large $a_2,x_2\geq p_2$,\\$\{x_2,y_2\}$}} (3)
    (3) edge [in = 30, out =-150] node [right] {\large $b_1,\top,\{\}$} (2)
    (3) edge [in = 90, out =-30] node [below] {}(7)
    (4) edge [in = 150, out =30] node [above] {\makecell{\large $a_2,x_2\geq p_2\wedge$\\$y_1<D_1-D_2$,\\$\{x_2,y_2\}$} }(8)
    (5) edge [in = -90, out =60] node [above] { }(8)
    (4) edge [in = 120, out =-60] node [below] {\large $a_2,x_2\geq p_2,\{x_2,y_2\}$} (6)
    (5) edge [in = -90, out =45] node [below right] {} (6)
    (5) edge [in = 150, out =30] node [below] {\makecell{\large $a_1,x_1\geq p_1\wedge$ \\$ y_2 < D_2 -D_1,\{x_1,y_1\}$}} (7)
    (6) edge[in = 0, out = 180] node [above]{\large $b_2,\top,\{\}$} (3)
    (8) edge[in = -30, out = -150] node [below right]{\large $b_2,\top,\{\}$} (4)
    (5) edge [in = -30, out = 150] node [below] {\large $b_2,\top,\{\}$} (2)
    (7) edge[in = -30, out = -150] node [below]{\large $b_1,\top,\{\}$} (5)
    (4) edge [in = 30, out = -150] node [above] {\large $b_2,\top,\{\}$} (1);
    \end{tikzpicture}
    }
    \caption{The timed automaton of the scheduling example.}
    \label{fig:Mot}
\end{figure}
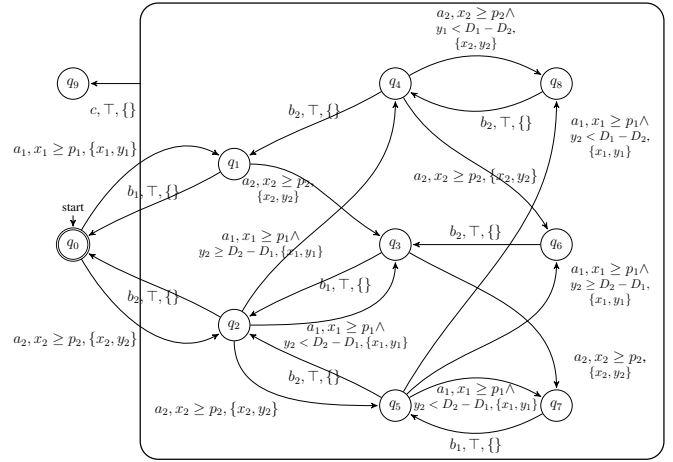

\begin{figure}[!t]
\centering
\resizebox{0.85\linewidth}{!}{
\begin{tikzpicture}[scale=0.8,>=stealth]
\draw[thick, fill=white!30] (0,1) rectangle (2,2) node[midway] {\large $\tau_1^1$};
\draw[thick, fill=white!30] (2,1) rectangle (2.8,2) node[midway] {\large $\tau_2^1$};
\draw[thick, fill=white!30] (2.8,1) rectangle (4.8,2) node[midway] {\large $\tau_1^2$};
\draw[thick, fill=white!30] (4.8,1) rectangle (6.3,2) node[midway] {\large $\tau_2^1$};
\draw[thick, fill=white!30] (6.3,1) rectangle (7.5,2) node[midway] {\large $\tau_2^2$};
\draw[thick, fill=white!30] (7.5,1) rectangle (9.5,2) node[midway] {\large $\tau_1^3$};

\draw[thick, fill=white!30] (0,0) rectangle (2,-1) node[midway] {\large $q_3$};
\draw[thick, fill=white!30] (2,0) rectangle (2.8,-1) node[midway] {\large $q_2$};
\draw[thick, fill=white!30] (2.8,0) rectangle (4,-1) node[midway] {\large $q_3$};
\draw[thick, fill=white!30] (4,0) rectangle (4.8,-1) node[midway] {\large $q_7$};
\draw[thick, fill=white!30] (4.8,0) rectangle (6.3,-1) node[midway] {\large $q_5$};
\draw[thick, fill=white!30] (6.3,0) rectangle (7.5,-1) node[midway] {\large $q_2$};
\draw[thick, fill=white!30] (7.5,0) rectangle (9.5,-1) node[midway] {\large $q_3$};

\draw[->] (0,-1.5) -- (10,-1.5) node[above] {Time};

\foreach \x in {0}
    \draw (\x cm,-1.2) -- (\x cm,-1.8) node[left] {$\x$};

\draw[thick, ->] (0,-2.2) -- (0,-1.5) node[below=0.5] {\large $a_1,a_2$};
\draw[thick, ->] (2,-2.2) -- (2,-1.5) node[below=0.4] {\large $b_1$};
\draw[thick, ->] (2.8,-2.2) -- (2.8,-1.5) node[below=0.5] {\large $a_1$};
\draw[thick, ->] (4,-2.2) -- (4,-1.5) node[below=0.5] {\large $a_2$};
\draw[thick, ->] (4.8,-2.2) -- (4.8,-1.5) node[below=0.4] {\large $b_1$};
\draw[thick, ->] (6.3,-2.2) -- (6.3,-1.5) node[below=0.4] {\large $b_2$};
\draw[thick, ->] (7.5,-2.2) -- (7.5,-1.5) node[below=0.5] {\large $a_1$};
\draw[thick, ->] (9.5,-2.2) -- (9.5,-1.5) node[below=0.4] {\large $b_1$};

\node[above right] at (-0.2,2) {Executing jobs on the processor};
\node[above right] at (-0.2,0) {Corresponding states of TA in Fig.~\ref{fig:Mot}};

\end{tikzpicture}
}
\caption{A possible schedule scenario over time with the executing jobs on the processor and the corresponding states of TA in Fig.~\ref{fig:Mot}. $\tau_i^j$ denotes the $j$-th job of the task $\tau_i$.}
\label{fig:schedule}
\end{figure}
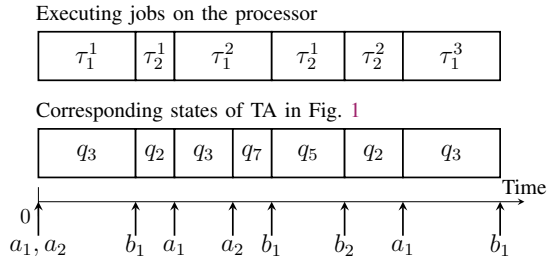

For instance, let us consider a case of a preemptive EDF scheduling problem with 2 tasks as follows. 
Two sporadic tasks $\tau_1, \tau_2$ run in parallel on one processor, and they have unknown minimum interarrival times $ p_i$, deadlines $D_i$ and controllable worst case execution time $C_i$, where $i=1,2$. At the same time, some constraints of the parameters are known: $p_1\leq p_2$, $D_1=p_1$, and $p_2 < D_2 \leq 2p_2$. Based on the known information, the system of task model with the preemptive  EDF strategy can be modeled as a TA shown in Fig.~\ref{fig:Mot}.
With preemptive EDF scheduling, the job currently executing on the processor is the first one in the priority queue determined by deadlines. Since $D_1=p_1$, there is at most 1 job of $\tau_1$  before a deadline is missed. Likewise, there are at most 2 jobs of $\tau_2$ in the queue before deadline misses. 
In Fig.~\ref{fig:Mot}, state $q_0$ stands for system idling, $q_1$ (resp. $q_2$) for only  $\tau_1$ (resp. $\tau_2$) in the queue, $q_3$ for 
$\tau_1, \tau_2$ in the queue, 
$q_4$ for 
$\tau_2,\tau_1$ in the queue,
$q_5$ for 
$\tau_2,\tau_2$ in the queue,
$q_6$ for 
$\tau_2,\tau_1,\tau_2$ in the queue,
$q_7$ for 
$\tau_1,\tau_2,\tau_2$ in the queue, and $q_8$ for 
$\tau_2,\tau_2,\tau_1$ in the queue. 
And $q_9$ is the state for missing deadlines.
Event $a_i$ stands for the arrival of jobs of $\tau_i$, and $b_i$ for the completion of  $\tau_i$, where $i=1,2$. Event $c$ stands for the claim of errors when any deadline is missed. Clock $x_i$ is used to record the separation time of jobs of $\tau_i$, and clock $y_i$ is used to record the time from the latest jobs of $\tau_i$ released up to now, where $i=1,2$.  A transition of the TA can be taken when the event happens and the value of the clocks satisfies the inequalities (guards), and after taking the transition, some clocks are reset to 0. For instance, the transition from $q_1$ to $q_3$ means that the next job of $\tau_2$ arrives and the current job of $\tau_1$ is not yet completed ($x_2 \geq p_2$), then the clocks $x_2$, $y_2$ for $\tau_2$ are reset to 0, and the system moves to $q_3$. And for any state except $q_0, q_9$, the transition to $q_9$ is taken if a deadline miss is claimed. 

Fig.~\ref{fig:schedule} illustrates a possible schedule scenario over time with the executing jobs and corresponding states of the TA in Fig.~\ref{fig:Mot}. $\tau_i^j$ denotes the $j$-th job of the task $\tau_i$. At the beginning, suppose each task has a job in the queue. Since the priority of $\tau_1$ is higher than that of $\tau_2$ (as $D_1 < D_2$), the TA is at State $q_3$ and the processor executes $\tau_1^1$ until finishing it. After that, the job $\tau_2^1$ is executed for a while, then the second job $\tau_1^2$ of the task $\tau_1$ comes and preempts the execution of $\tau_2^1$. When the second job of $\tau_2$ comes, the TA jumps from $q_3$ to $q_7$. After finishing the second job of $\tau_1$, the TA jumps from $q_7$ to $q_5$ and the processor completes the execution of $\tau_2^1$. After that, the processor executes the second job of $\tau_2$ for a while, before the third job of $\tau_1$ comes to preempt its execution. 

To determine the parameters, we can learn a formal model, say a TA or a TRE, from the observed behaviors. After that, we can determine the condition for the execution times $C_1$ and $C_2$ so that the system scheduling is feasible. We present it as a case study in Section~\ref{sbsc:case_study}.

\section{Preliminaries} \label{sc:preliminary}

In this section, we review the basic notions and results on timed regular expressions. Please refer to~\cite{AsarinCM02} for a comprehensive introduction. 


Let $\mathbb{R}_{\geq 0}$ and $\mathbb{N}$ represent non-negative reals and natural numbers, respectively. 
\emph{An integer-bounded interval} $I$ is either $[l,u]$, $(l,u]$, $[l,u)$, $(l,u)$, $[l,\infty)$ or $(l,\infty)$, where $l,u\in\mathbb{N}$ such that $l\leq u$. Let $\Sigma$ be a fixed finite alphabet. In practice, an event $\sigma\in\Sigma$ denotes an action of a system under consideration. 



A \emph{timed word} is a finite sequence $\omega=(\sigma_1,t_1)(\sigma_2,t_2)\cdots\\(\sigma_n,t_n)$ $ \in (\Sigma\times\mathbb{R}_{\geq 0})^*$, where $t_i$ represents the delay time before taking action $\sigma_i$ for all $1\leq i\leq n$.
Particularly, the special symbol $\varepsilon$ represents the \emph{empty word}. For a timed word $\omega$, we define two functions $\lambda: (\Sigma\times\mathbb{R}_{\geq 0})^*\rightarrow\mathbb{R}_{\geq 0}^*$ and $\mu: (\Sigma\times\mathbb{R}_{\geq 0})^*\rightarrow\Sigma^*$ to get the sequence of delay times (i.e., $\lambda(\omega)=t_1,t_2,\dots,t_n\in\mathbb{R}_{\geq 0}^*$) and the sequence of events (i.e., $\mu(\omega)=\sigma_1,\sigma_2,\dots,\sigma_n\in\Sigma^*$), respectively. For instance, let $\omega=(a,0.2)(b,0)(a,6)$, then we have $\lambda(\omega)=0.2,0,6$ and $\mu(\omega)=aba$. A \emph{timed language} is a set of timed words.


According to~\cite{AsarinCM02}, the full set of timed regular expressions, named generalized extended timed regular expressions ($\mathcal{GEE}(\Sigma)$), admits the following operations: \emph{concatenation} ($\cdot$), \emph{union} ($\vee$), \emph{time restriction} ($\langle\,\rangle_{I}$), \emph{Kleene star} ($^*$), \emph{absorbing concatenation} ($\circ$), \emph{absorbing iteration} ($^\circledast$), \emph{intersection} ($\wedge$), and \emph{renaming} ($\theta$). It is proved that $\mathcal{GEE}(\Sigma)$ has the same expressive power as timed automata~\cite{Alur94}. Renaming function is needed to show the expressiveness of $\mathcal{GEE}(\Sigma)$ but not easy to infer from system behaviors. In this paper, we consider the standard timed regular expressions ($\cdot$, $^*$, $\vee$ and $\langle\,\rangle_{I}$) and the extended timed regular expressions ($\circ$ and $^\circledast$ additionally). Since the extended timed regular expressions ($\mathcal{EE}(\Sigma)$) have the same expressive power as standard timed regular expressions ($\mathcal{E}(\Sigma)$),
we thus use the following definition of standard timed regular expressions directly.

\begin{mydefinition}[Timed regular expressions] \label{def:tre}
The syntax of timed regular expressions (TRE) over $\Sigma$ is as follows:
\[
\varphi := \varepsilon \;\vert\; \underline{\sigma} \;\vert\; \langle\varphi\rangle_{I} \;\vert\; \varphi\cdot\varphi \;\vert\; \varphi\vee\varphi \;\vert\; \varphi^*
\vspace{-0.05cm}
\]
where $\underline{\sigma}$ for every event $\sigma\in\Sigma$ and $I$ is an integer-bounded interval, $\cdot$, $\vee$, and $*$ stand for concatenation, disjunction, and Kleene star, respectively.
The semantics of TRE $\llbracket\;\rrbracket:\mathcal{E}(\Sigma)\rightarrow 2^{(\Sigma\times\mathbb{R}_{\geq 0})^*}$, which associates a TRE with a set of timed words, is inductively defined as follows:
\begin{gather*}\label{eq}
\semantic{\varepsilon}  = \{\varepsilon\}, \quad
\semantic{\underline{\sigma}} = \{(\sigma,t):t\in\mathbb{R}_{\geq 0}\}, \quad
\semantic{\langle\varphi\rangle_{I}} = \semantic{\varphi} \cap \\\{\omega:\sum\lambda(\omega)\in I\}, \quad
\semantic{\varphi_1\cdot\varphi_2} = \semantic{\varphi_1} \cdot \semantic{\varphi_2}, \quad \\
\semantic{\varphi_1\vee\varphi_2}  =  \semantic{\varphi_1} \cup \semantic{\varphi_2}, \quad 
\semantic{\varphi^*}  = \bigcup\nolimits_{i=0}^{\infty}(\llbracket\underbrace{\varphi\cdot \ldots \cdot\varphi}_\text{$i$ times}\rrbracket)
\end{gather*}
\end{mydefinition}

A TRE represents a timed language. Two timed words $\omega_1$ and $\omega_2$ can be \emph{distinguished} by a TRE $\varphi$ if $(\omega_1\in \semantic{\varphi} and$ $   \omega_2\not\in \semantic{\varphi})$ or $(\omega_1 \not\in \semantic{\varphi} and$ $   \omega_2\in \semantic{\varphi})$. To determine whether $\omega\in\semantic{\varphi}$ is called as the \emph{membership decision}, which can be done by a standard method~\cite{AsarinCM02}. 
The meaning of $\underline{\sigma}$ for every event $\sigma\in\Sigma$ represents the occurrence of $\sigma$ after delaying an arbitrary time. The \emph{time restriction (constraint)} $\langle\varphi\rangle_{I}$ limits the total time passages in the interval $I$ over timed words $\omega$ in $\llbracket\varphi\rrbracket$. It is obvious that $\semantic{\tr{\varepsilon}_{I}}=\{\varepsilon\}$. Other operations $\vee$, $\cdot$ and $^*$ satisfy the standard Kleene algebra and we use the following conventions $\sigma=\langle\underline{\sigma}\rangle_{[0,0]}$, $\varphi^0=\varepsilon$, $\varphi^+=\varphi\cdot\varphi^*$, and $\varphi^{i+1}=\varphi^{i}\cdot\varphi=\varphi\cdot\varphi^{i}$. We will omit the concatenation symbol ($\cdot$) if the context is clear. Therefore, compared to regular expressions, the most important new feature here is the time restriction. 
The following examples further illustrate it.

\noindent\resizebox{\linewidth}{!}{
\begin{minipage}{\linewidth}
\vspace{-0.1cm}
\begin{align*}
    \varphi_1  & = \tr{\underline{\sigma}}_{[0,0]}  \\ \semantic{\varphi_1}  & = \cbrackets{(\sigma,0)} \\[0.1cm]
    \varphi_2  & = \tr{\underline{\sigma}}_{[3,7]}  \\ \semantic{\varphi_2}  & = \cbrackets{(\sigma,t):t\in[3,7]} \\[0.1cm]
    \varphi_3  & = \tr{\underline{\sigma_1}}_{[1,3]} \cdot \tr{\underline{\sigma_2}}_{[2,4]} \\ \semantic{\varphi_3}  & = \cbrackets{(\sigma_1,t_1)(\sigma_2,t_2):t_1\in[1,3]\wedge t_2\in[2,4]} \\[0.1cm]
    \varphi_4  & = \tr{\tr{\underline{\sigma_1}}_{[1,3]} \cdot \underline{\sigma_2}}_{[2,4]} \\ \semantic{\varphi_4}  & = \cbrackets{(\sigma_1,t_1)(\sigma_2,t_2):t_1\in[1,3]\wedge t_1 + t_2\in[2,4]} \\[0.1cm]
    \varphi_5  & = \tr{\tr{\underline{\sigma_1}}_{[1,3]} \cdot \tr{\underline{\sigma_2}}_{[2,4]}}_{[5,6]} \\ \semantic{\varphi_5}  & = \cbrackets{(\sigma_1,t_1)(\sigma_2,t_2):t_1\in[1,3]\wedge t_2\in[2,4]\wedge t_1 + t_2 \in[5,6]} \\[0.1cm]
    \varphi_6  & = \tr{\underline{\sigma}^*}_{[3,7]} \\ \semantic{\varphi_6}  & = \cbrackets{(\sigma,t_1)(\sigma,t_2)\cdots(\sigma,t_n):n\in\mathbb{N}\wedge\sum\nolimits_{i=1}^{n}{t_i}\in[3,7]} \\[0.1cm]
    \varphi_7  & = \pbrackets{\tr{\underline{\sigma}}_{[3,7]}}^* \\ \semantic{\varphi_7} & = \cbrackets{(\sigma,t_1)(\sigma,t_2)\cdots(\sigma,t_n): n\in\mathbb{N}\wedge \bigwedge\nolimits_{i=1}^{n}{t_i\in[3,7]}}
\end{align*}
\end{minipage}
} 

We reuse the function $\mu$ to get the corresponding regular expression (RE) $\mu(\varphi)$ by erasing the time restrictions in a TRE $\varphi$. For instances, $\mu(\varphi_1)=\mu(\varphi_2)=\sigma$, $\mu(\varphi_3)=\mu(\varphi_4)=\mu(\varphi_5)=\sigma_1\cdot\sigma_2$, and $\mu(\varphi_6)=\mu(\varphi_7)=\sigma^*$. 

A \emph{TRE syntax tree} (see the example in Fig.~\ref{fig:encoding} in Section \ref{sc:syn_pro}) allows to represent a TRE with a tree structure, where internal nodes are labeled by operations and leaf nodes are labeled by actions in $\Sigma$. An internal node labeled with a unary operation ($^*$) has exactly one child. An internal node labeled with a binary operation ($\cdot$ or $\vee$) has exactly two children. Each node is equipped with a time restriction $\langle\rangle_{I}$ as an inherent property.



\section{Problem Formulation} \label{sc:problems}
In this section, we first define the TRE synthesis problems of our interest. Then we 
discuss several distinctions between synthesizing TRE and traditional RE.

\subsection{Problem Statements} \label{sbsc:problem_statements}
In general, our purpose is to synthesize a TRE $\varphi$ from a given finite set of timed words $\Omega\subseteq(\Sigma\times\mathbb{R}_{\geq 0})^*$. 
One particular setting is that the set of examples is given by a pair $\Omega=(\Omega_{+},\Omega_{-})$, where $\Omega_{+}$ contains the positive examples and $\Omega_{-}$ contains the negative ones.


\oomit{
\begin{remark}
    In the case of synthesizing RE with only positive examples, there is always a naive solution $\varphi=\bigvee_{\omega\in\Omega}{\underline{\omega}}$, where $\underline{\omega}$ is the simple RE of an example $\omega$. And if negative examples are considered, one just needs to check if positive and negative examples clash, meaning the two sets have some common element. In turn, if the clash occurs, there is no RE to be synthesized. Likewise, in timed cases we can imagine a "naive solution" based on positive examples when no clash occurs, but the question is how to define "clash" in timed case. 
\end{remark}
}

\begin{myproblem}[Synthesis problem]\label{prob:syn_p_n}
Given a finite set of timed words $\Omega=(\Omega_{+},\Omega_{-})$, to synthesize a TRE $\varphi$ such that for all $\omega \in \Omega_{+}, \omega\in\semantic{\varphi} and$ for all $\omega \in \Omega_{-}, \omega\not\in\semantic{\varphi}$, or to claim that there does not exist such TRE. In the first case, we say $\phi$ recognizes $\Omega$. 
\end{myproblem}


To specify ideal expressions, we can introduce measures in TRE. The motivation is that for Problem~\ref{prob:syn_p_n} we need some measure to determine an interesting expression since there may be infinitely many such expressions that recognize a given set $\Omega$. Formally, a measure $\mathcal{M}:\mathcal{E}(\Sigma)\rightarrow \mathbb{R}_{\geq 0}\cup\{\infty\}$ assigns a real value to every expression so that we can compare them.

However, there does not appear to be any uniform agreement on how to measure the size of even untimed regular expressions on the alphabet $\Sigma$. An obvious measure is the ordinary length, i.e., the total number of symbols including parentheses. Another measure is the formula length based on the reverse Polish form. Evidently, reverse polish length is the same as the number of nodes in the syntax tree for the expression. 

For TRE, we are interested in the number of nodes in the syntax tree with considering time restriction $\tr{\,}_{I}$ as a property for each node. We say the \emph{length} of a TRE is the number of nodes in the corresponding syntax tree.

\begin{myproblem}[Minimal synthesis problem] \label{prob:min_syn}
Given a finite set of timed words $\Omega$, synthesize a TRE $\varphi$ with minimal length recognizing  $\Omega$, or claim that there is no such TRE.
\end{myproblem}

\subsection{Differences from synthesizing regular expressions} \label{sbsc:differences_from_RE}

\myparagraph{Synthesizing time restrictions} We not only synthesize untimed expressions but also time constraints. Time constraints are not only for the letters in expressions but also for the constraints over the operations (subformulas). TRE containing time constraints only for letters is equivalent to real-time automata, which is a kind of timed automata with only one clock and resets it at every transition.

\myparagraph{Rejecting negative examples} An RE rejects a string over $\Sigma$ iff the string cannot match the expression. However, in our case, a TRE rejects a timed word over $\Sigma\times\mathbb{R}_{\geq 0}$ is additionally due to timing information. Therefore, given a negative example $\omega$ and a TRE $\varphi$, the string $\mu(\omega)$ may match the corresponding untimed RE. It brings the obstacle of utilizing the current pruning methods of synthesizing regular expressions to the synthesis of TRE.

\myparagraph{Pruning structure} Given two RE $r_1 = (\underline{a}\cdot \underline{b})^*$ and $r_2 = (\underline{a}\cdot \underline{b})^*\cdot(\underline{a}\cdot \underline{b})^*$, we have that $r_1 = r_2$ and $r_1$ is smaller than $r_2$ in length. If $r_1$ does not 
recognize 
the examples, then we do not need to take $r_2$ into consideration anymore and thus skip it in the searching process. However, this is not the case for synthesizing TRE. For instance, $\varphi_1=(\tr{\underline{a}\cdot \underline{b}}_{[1,3]})^*$ and $\varphi_2=(\tr{\underline{a}\cdot \underline{b}}_{[1,3]})^*\cdot (\tr{\underline{a}\cdot \underline{b}}_{[4,6]})^*$, we have $\semantic{\varphi_1}\subset\semantic{\varphi_2}$. Even if we cannot synthesize ideal intervals on the untimed expression $(ab)^*$, we cannot skip the equivalent expression $(ab)^*\cdot(ab)^*$.

\myparagraph{Equivalent structure} For example, the RE structure $(\underline{a}\cdot \underline{b})\cdot \underline{c}$ is equivalent to $\underline{a}\cdot(\underline{b}\cdot \underline{c})$. But they are not equivalent anymore as we equip every concatenation $\cdot$ with a time restriction, since the time restrictions of the first and the second concatenation constraint the delay times of timed words differently in the two structures. So we cannot skip one of the structures if we have already checked the other. 

\myparagraph{Limitation} For synthesizing REs from positive and negative examples, if we assume that the two sets are finite and disjoint, then we can always get a valid expression of which the language is not empty. However, for the case of synthesizing TREs, even if the two sets are finite and disjoint, we may claim that there is no such TRE recognizing the examples. One reason is that we can not generate valid integer-bounded intervals. Another reason is that we need to use other operations, e.g., intersection, if we want to represent some timed languages by TREs. For example, a $\mathcal{GEE}$ expression $(\tr{a\cdot b}_{[1,1]}\cdot c)\wedge(a\cdot\tr{b\cdot c}_{[1,1]})$ denoting the set $\{(a,t_1)(b,t_2)(c,t_3):(t_1 + t_2=1)\wedge(t_2+t_3=1)\}$ can not be expressed without intersection. 

\section{Decidability of TRE Synthesis Problems} \label{sc:decidability}


For Problem~\ref{prob:syn_p_n}, one can imagine that there exists infinite many of TREs accepting all positive examples, for instance, since $\Sigma$ is finite,  $\varphi=\tr{\tr{\tr{\underline{\sigma_1}}_{[0,\infty)}\vee_ {[0,\infty)}\tr{\underline{\sigma_2}}_{[0,\infty)}\vee_ {[0,\infty)}\cdots \vee_ {[0,\infty)}\tr{\underline{\sigma_n}}_{[0,\infty)}}^*}_{[0,\infty)}$, where $\Sigma=\{\sigma_1,\sigma_2,\cdots,\sigma_n\}$ is a ``na\"ive'' solution TRE accepting all timed words $w\in(\Sigma\times\mathbb{R}_{\geq 0})^*$. However, when considering negative examples, the learnt TRE should reject all negative examples because of at least one of the following two reasons: untimed sequences are unmatched or the delay time information is not acceptable. Considering the second condition, given a positive example $\omega$ and a negative example $\omega'$ such that $\mu(\omega)=\mu(\omega')$, the time restrictions in a candidate TRE should be able to reject $\lambda(\omega')$. 
Therefore, we introduce the \emph{simple timed regular expression} to decide whether a TRE is able to distinguish any pair of positive and negative examples.






\begin{mydefinition}[Simple elementary language~\cite{Waga23}] \label{def:sel}
Given a timed word $\omega=(\sigma_1,t_1)(\sigma_2,t_2)\cdots(\sigma_n,t_n)$, the corresponding \emph{simple elementary language} (sEL) is a timed language $\mathcal{SE}({\omega}) =  \{(\sigma_1,t_1')(\sigma_2,t_2')\cdots(\sigma_n,t_n'):t_1',t_2',\cdots,t_n' \models \Lambda(\omega)\}$, where $\Lambda(\omega)=\bigwedge\Theta(\omega)$ and $\lambda(\omega)$ satisfies $\Lambda(\omega)$, i.e., $\lambda(\omega) \models \Lambda(\omega)$. $\Theta(\omega)$ is the finite set of tightest integer-bounded time constraints for the sums of all possible consecutive sequences of delay times of $w$. i.e. $\forall 1\le j\le k\le n$, let $d$ be the integer part of $\sum\nolimits_{i=j}^{k}{t_i}$ and $e$ be the fractional part, if $e>0$ then inequality $d < \sum\nolimits_{i=j}^{k}{t_i}< d+1 $ is in $ \Theta(\omega)$, otherwise equality $\sum\nolimits_{i=j}^{k}{t_i} = d $ is in $ \Theta(\omega)$.
\end{mydefinition}

\begin{myexample}
Considering timed words $\omega_1=(a,1.2)(a,2.2)$, $\omega_2=(a,1.2)(a,2.6)$, $\omega_3=(a,1.5)(a,2.6)$, we have $\mathcal{SE}(\omega_1)=\mathcal{SE}(\omega_2)=\semantic{\tr{\tr{\underline{a}}_{(1,2)} \cdot \tr{\underline{a}}_{(2,3)}}_{(3,4)}}$, and $\mathcal{SE}(\omega_3)=\semantic{\tr{\tr{\underline{a}}_{(1,2)} \cdot \tr{\underline{a}}_{(2,3)}}_{(4,5)}}$
\end{myexample}

\begin{mylemma}\label{lemma:sEL}
Given two timed words $\omega_1$ and $\omega_2$, if $\mathcal{SE}(\omega_1)=\mathcal{SE}(\omega_2) $, then  $\omega_1$ and $\omega_2$ cannot be distinguished by any TRE. 
\end{mylemma}

However, not all elementary languages necessarily have a corresponding TRE whose semantics are equal to the elementary language. For instance, given $\omega_1=(a,1.4)(a,1.4)(a,1.4)$ we have $\mathcal{SE}(\omega_1)=\{(a,\tau_1)(a,\tau_2)(a,\tau_3):\tau_1,\tau_2,\tau_3 \in (1,2),$
$\tau_1+\tau_2,\tau_2+\tau_3 \in (2,3), \tau_1+\tau_2+\tau_3 \in (4,5)\}$. There does not exist a TRE $\varphi$ s.t. $\semantic{\varphi}=\mathcal{SE}(\omega_1)$, since $\tau_2$ appears in both constraints of $\tau_1 + \tau_2$ and $\tau_2+\tau_3$, which is not allowed according to Definition~\ref{def:tre}.
Hence, we loose the requirement of sEL to define simple timed regular expressions as follows.

\begin{mydefinition}[Simple timed regular expressions] \label{def:stre}
A simple timed regular expression (sTRE) $\varphi$ contains only two operations $
\cdot$ and $\tr{}_{I}$, where all time restrictions $I$ are in the form of $(d,d+1)$ or $[d,d]$, and $d\in\mathbb{N}$.
\end{mydefinition}

Given a timed word $\omega$, we denote $\Phi(\omega)$ the finite set of its corresponding sTRE, and write $\varphi_\omega$ for one of the corresponding sTRE, i.e., $\varphi_\omega\in\Phi(\omega)$. 

\begin{mylemma} \label{lemma:num_sTRE}
Given a timed word $\omega$, the number of its corresponding sTRE $\varphi$ (i.e., $\omega\in\semantic{\varphi}$) is at most $2^{(n^2+n)/2}$, where $n=\lvert \omega \rvert$. \notag
\end{mylemma}
The following lemma shows that the simple elementary language of a timed word is identical to the language intersection of all its corresponding sTRE. 


\begin{mylemma} \label{lemma:relation_sEL_sTRE}
Given a timed word $\omega$, $\mathcal{SE}(\omega)=\bigcap_{\varphi_\omega\in\Phi(\omega)}{\semantic{\varphi_\omega}}$.
\end{mylemma}

\begin{mydefinition}[Obscuration] \label{def:falsifiability}
A timed word $\omega$ is obscured by a set of timed words $\Omega$, iff $\forall \varphi_\omega \in \Phi(\omega), \exists \omega'\in{\Omega}, \omega' \in \semantic{\varphi_\omega}$. A timed word $\omega$ is 
not obscured by a set of timed words $\Omega$, iff $\exists \varphi_\omega \in \Phi(\omega), \forall \omega'\in{\Omega}, \omega' \not\in \semantic{\varphi_\omega}$.  
\end{mydefinition}

\begin{myexample}
Consider $\omega_1=(a,1.5)(a,2.6)(a,1.5)$, $\omega_2=(a,1.2)(a,2.6)(a,1.5)$, $\omega_3=(a,1.5)(a,2.6)(a,1.2)$. Timed word $\omega_1$ is not obscured by $\{\omega_2\}$ or $\{\omega_3\}$, but is obscured by $\{\omega_2, \omega_3\}$. For instance, $\omega_1,\omega_3\in\semantic{\tr{\tr{\underline{a}\cdot \underline{a}}_{(4,5)}\cdot\underline{a}}_{(5,6)}}$, $\omega_2\not\in\semantic{\tr{\tr{\underline{a}\cdot \underline{a}}_{(4,5)}\cdot\underline{a}}_{(5,6)}}$, $\omega_1,\omega_2\in\semantic{\tr{\underline{a}\cdot\tr{\underline{a}\cdot \underline{a}}_{(4,5)}}_{(5,6)}}$, $\omega_3\not\in\semantic{\tr{\underline{a}\cdot\tr{\underline{a}\cdot \underline{a}}_{(4,5)}}_{(5,6)}}$.
\end{myexample}

\begin{mylemma}\label{lemma:fals_dcd}
Given a timed word $\omega$, the obscuration of it w.r.t a set of timed words $\Omega$ is decidable in at most $2^{(n^2+n)/2}m$ TRE membership decisions, where $n=\lvert \omega \rvert$  and $m=\lvert \Omega \rvert$.
\end{mylemma}

\oomit{
\begin{proof}
 From lemma \ref{lemma:num_sTRE}, it takes at most $2^{(n^2+n)/2}$ TRE acceptance queries to check if $\exists \varphi_\omega \in \Phi(\omega), \omega'\in{\Omega}$, $\omega' \in \semantic{\varphi_\omega}$. So it takes at most $2^{(n^2+n)/2}\cdot m$ TRE acceptance queries to check all $m$ timed words in $\Omega$.  
\end{proof}
}

\begin{mylemma}\label{lemma:stre_fals}
If a positive example $\omega \in \Omega_{+}$ is obscured by the set of negative examples $\Omega_{-}$, for all TRE $\phi$ such that $\omega\in\semantic{\phi}$, there exists $\omega'\in{\Omega_{-}}$ such that $\omega'\in\semantic{\phi}$.
\end{mylemma}
\oomit{
\begin{proof}
By parsing how $\omega$ is accepted by $\phi$ we yield $\lambda \models \Lambda^{\phi}({\omega})$, where $\Lambda^{\phi}({\omega})$ is the set of inequalities (and equalities) about $\lambda(\omega)$ in current parsing situation.
Then there exists a sTRE $\varphi$ of $\omega$, whose time constraints $\Lambda^{\varphi}({\omega})$ is of the same form of $\Lambda^{\phi}({\omega})$, with only differences on the numbers in the inequalities (and equalities) that come from the interval bounds in $\phi$ and $\varphi$. Suppose one of the inequalities in $\Lambda^{\phi}({\omega})$, the corresponding inequality in $\Lambda^{\varphi}({\omega})$ must be the same or more strict by the definition of sTRE. So $\semantic{\varphi}\subseteq\semantic{\phi}$.
Then by the definition of obscuration, $\exists \omega_{-}\in{\Omega_{-}}, \omega_{-}\in\semantic{\varphi}\subseteq\semantic{\phi}$.
\end{proof}
}
\oomit{
\begin{proof}
By parsing how $\omega$ is accepted by $\phi$ we yield $\lambda \models \Lambda^{\phi}({\omega})$, where $\Lambda^{\phi}({\omega})$ is the set of inequality in current parsing situation.
Then there exists a sTRE $\varphi$ of $\omega$, whose time constraints is of the same form of $\Lambda^{\phi}({\omega})$, with only differences on the interval bounds.
By definition, $\exists \omega_{-}\in{\Omega_{-}}, \omega_{-}\in\semantic{\phi}\subseteq\semantic{\phi}$.\qed  
\end{proof}
}

\begin{mytheorem} \label{thm:decidability}
Problem~\ref{prob:syn_p_n} has a solution TRE w.r.t a given sets $\Omega = (\Omega_+,\Omega_-)$ iff $\forall \omega_{+}\in{\Omega_{+}}$, $\omega_{+}$ is not obscured by $\Omega_-$. If so, one solution to Problem~\ref{prob:syn_p_n} is $\varphi=\bigvee_{\omega_+\in\Omega_+}\varphi_{\omega_+}$.
\end{mytheorem}
\oomit{
\begin{proof}
If $\forall \omega_{+}\in{\Omega_{+}}$ is not obscured by $\Omega$, then for each $\omega_+$, there exists a sTRE $\varphi\in\Phi(\omega_+)$ such that $\forall\omega_{-}\in{\Omega_{-}},\omega_-\not\in\semantic{\varphi}$, which we denote as $\varphi_t(\omega_+)$ here. $\varphi=\bigvee_{\omega_+\in \Omega_+}\varphi_t(\omega_+)$ is a solution of Problem \ref{prob:syn_p_n}. 

If there exists $\omega_{+}\in{\Omega_{+}}$ that is obscured by $\Omega_-$, and suppose $\varphi'$ is a solution of Problem \ref{prob:syn_p_n}, then $\omega_{+}\in\semantic{\varphi'}$. By Lemma~\ref{lemma:stre_fals}, there exists $\omega_{-}\in{\Omega_{-}}, \omega_{-}\in\semantic{\varphi'} $, which is a contradiction. 
\end{proof}
}

By Lemma \ref{lemma:fals_dcd} and Theorem~\ref{thm:decidability}, we have the conclusion that Problem~\ref{prob:syn_p_n} is decidable. In what follows, suppose that Problem~\ref{prob:syn_p_n} has a solution w.r.t a given set $\Omega$, we investigate Problem~\ref{prob:min_syn}, i.e., synthesizing a TRE with minimal length.



\section{TRE Synthesis}\label{sc:syn_pro}

\subsection{Overview of synthesis of TRE with minimal length} \label{sbsc:overview}
For the minimal synthesis problem (Problem~\ref{prob:min_syn}), the basic idea is that we enumerate TRE with increasing length and check if the current candidate recognizes the given set $\Omega$. However, unlike traditional RE, even for a finite alphabet, the number of $k$-length TRE is infinite, making it impossible to enumerate TRE in this na\"ive way.


Therefore, we introduce a structure, named \emph{parametric timed regular expression} ($p$TRE), equipped with \emph{parametric intervals} replacing of time restrictions, which is suitable for enumeration. 

\begin{mydefinition}[Parametric interval]
A parametric interval $\mathbb{I}$ is a 4-tuple $(lo, l, u, uo)$ where two integer variables $l,u$ denote the lower and upper endpoints of the interval, and two Boolean variables $lo,uo$ indicate whether the interval includes the start and end point, respectively. 
\end{mydefinition}

\begin{mydefinition}[Parametric timed regular expressions]
The syntax of parametric timed regular expressions ($p$TRE) over $\Sigma$ is given by the grammar:
\[
\phi := \hole \;\vert\; \varepsilon \;\vert\; \tr{\underline{\sigma}}_{\mathbb{I}} \;\vert\;  \tr{\phi\cdot\phi}_{\mathbb{I}} \;\vert\; \tr{\phi\vee\phi}_{\mathbb{I}} \;\vert\; \tr{\phi^*}_{\mathbb{I}} 
\]
where $\hole$ is a placeholder for further enumeration of $p$TRE that can be substituted by expression $\phi$ s.t. $\phi\neq\hole$.
\end{mydefinition}

Particularly, a $p$TRE is termed a \emph{closed} $p$TRE if it contains no $\hole$. A $p$TRE is termed an \emph{edge} $p$TRE if it contains only one $\hole$.

If all parametric intervals $\mathbb{I}$ in a closed $p$TRE $\phi$ are determined, we say the resulting TRE $\varphi$ is an instance of $\phi$, denoted as $\varphi\in[\phi]$. Therefore, a $p$TRE $\phi$ has infinitely many instances. The \emph{length} of $p$TRE is defined as the number of nodes in the syntax tree. 

We denote a specific instance of a closed $p$TRE $\phi$ as $max(\phi)$, where all parametric intervals in $\phi$ are set to $[0,\infty)$. A time word $\omega$ is \emph{accepted} by a closed $p$TRE $\phi$ iff $\omega$ is accepted by $max(\phi)$.


\begin{mylemma}\label{lemma:ptre_finite}
The number of $p$TRE with length $k\in\mathbb{N}$ is finite.
\end{mylemma}
\oomit{
\begin{proof}
Starting with $k=1$, the number of $p$TRE is obviously finite. 

If the number of $p$TRE is finite at $k\leq n$, then at $k=n+1$, any $(n+1)$-length $p$TRE can be generated by applying $
\phi := \tr{\phi^*}_{\mathbb{I}} $ on a $n$-length $p$TRE or by applying $\phi := \tr{\phi\cdot\phi}_{\mathbb{I}} \;\vert\; \tr{\phi\vee\phi}_{\mathbb{I}}$ on a $(n-1)$-length $p$TRE. The number of $\phi$ in a $n$-length (or $(n-1)$-length) $p$TRE is finite, and applying $\phi := \tr{\phi^*}_{\mathbb{I}} $ (or $\phi := \tr{\phi\cdot\phi}_{\mathbb{I}} \;\vert\; \tr{\phi\vee\phi}_{\mathbb{I}}$) on a particular $\phi$ yields 1 (or 2) particular $p$TRE. So the number of $(n+1)$-length $p$TRE finite. By induction, $\forall k \in \mathbb{N}$, the number of $k$-length $p$TRE is finite.
\end{proof}
}

Given a set of positive examples $\Omega_{+}$, we write $p$TRE$_+$ for a closed $p$TRE accepting all positive examples $\omega_+\in\Omega_+$. By Lemma~\ref{lemma:ptre_finite}, the number of $p$TRE$_+$ with $k$ length is also finite.

\begin{algorithm}[!t]\label{algorithm: synthesis_overview}
	\caption{Synthesizing a TRE with minimal length}
	\SetKwInOut{Input}{input}
	\SetKwInOut{Output}{output}
	\Input{A sample set $\Omega=(\Omega_+,\Omega_-)$.}
	\Output{A minimal TRE $\varphi$ recognizing $\Omega$.}
        $k \gets 1$\; 
	\While{$\top$}{
            \ForEach{$\phi_+ \gets$ \text{enumerate}($k, \Omega_+$)}{\label{alg1_line:enumerate}
                $\xi \gets$ encode($\phi_+,\Omega_-$) \label{alg1_line:encode} 
                \tcp*{Encode $k$-length $p$TRE$_+$ $\phi_+$ once get it.} 
                $\mathit{flag}, \varphi \gets$ SMTsolver($\xi,\phi_+$)\;
                \If{$\mathit{flag}=\top$}{ \label{alg1_line:solution}
                \Return $\varphi$\;
                }
            }
            $k\gets k+1$; \label{alg1_line:len_increase}
        }
\end{algorithm}

Now, instead of enumerating TRE, we can enumerate $p$TRE$_+$. For each $p$TRE$_+$ $\phi_+$, since it already accepts all positive examples, we encode the requirement that $\phi_+$ rejects all negative examples $\Omega_-$ into an SMT formula. If a solution to the formula exists, we obtain instances for the parametric intervals inside and derive a TRE $\varphi$ recognizing $\Omega$. Since we systematically enumerate $p$TRE$_+$ with increasing length, it is ensured that the resulting TRE is a minimal one. Algorithm~\ref{algorithm: synthesis_overview} summarizes the synthesis process in a nutshell. Note that there may be a lot of $p$TRE$_+$ with a length $k$, we encode each $p$TRE$_+$ with $\Omega_-$ into an SMT formula $\xi$ greedily once we get one (Line~\ref{alg1_line:enumerate} and Line~\ref{alg1_line:encode}) instead of after collecting all of $k$-length $p$TRE$_+$. If there is a solution to $\xi$ (Line~\ref{alg1_line:solution}), we get a minimal TRE $\varphi$ recognizing $\Omega$. If there is no solution with $k$ length, we start to find $p$TRE$_+$ with length $k+1$ (Line~\ref{alg1_line:len_increase}). 

Before proving the termination and correctness of our method, we present the enumerating details and several pruning techniques utilized (Section~\ref{sc:enum_and_prune}), and then show the details of the encoding (Section~\ref{sc:time_cons}).

\subsection{Enumerating and pruning $p$TRE}\label{sc:enum_and_prune}


The following lemma claims that searching for minimal TRE (and thus the enumeration of $p$TRE) does not need to consider $\varepsilon$, since it is obvious that a $p$TRE with $\varepsilon$ has the same semantics as a shorter $p$TRE without $\varepsilon$.

\begin{mylemma}\label{lemma:no_epsilon}
$\varepsilon$ never appears in a minimal TRE w.r.t a nonempty set $\Omega$.
\end{mylemma}

Therefore, starting with a placeholder $\hole$, we consider the following substitution rule in the enumeration.

\begin{center}
    \AxiomC{}
    \LeftLabel{$1.$\,}
    \RightLabel{$,\sigma\in\Sigma$}
    \UnaryInfC{$\hole\rightarrow \tr{\underline{\sigma}}_{\mathbb{I}}$}
    \DisplayProof
    \hskip 1.5em
    \AxiomC{}
    \LeftLabel{$2.$\,}
    \UnaryInfC{$\hole\rightarrow \tr{\hole^*}_{\mathbb{I}}$}
    \DisplayProof
    \hskip 1.5em\\[0.2cm]
    \AxiomC{}
    \LeftLabel{$3.$\,}
    \UnaryInfC{$\hole\rightarrow \tr{\hole\cdot\hole}_{\mathbb{I}}$}
    \DisplayProof
    \hskip 1.5em
    \AxiomC{}
    \LeftLabel{$4.$\,}
    \UnaryInfC{$\hole\rightarrow \tr{\hole\vee\hole}_{\mathbb{I}}$}
    \DisplayProof
\end{center}


\begin{algorithm}[t]\label{algorithm: trivial}
	\caption{Trivial Enumeration of $p$TRE$_+$}
	\SetKwInOut{Input}{input}
	\SetKwInOut{Output}{output}
	\Input{A required length $k\in\mathbb{N}$; the positive samples $\Omega_+$}
	\Output{The set of $k$-length $p$TRE$_+$ $S_+$.}
	  $\mathit{rules} \gets \{2,3,4\}$ \;
        $S_k \gets$ substitute($\mathit{rules}$,$\hole$,$k$)\;
        \ForEach{$\phi \in S_k$}{
            $\mathit{rules}\gets \{1\}$\;
            $S_k' \gets$ substitute($\mathit{rules}$,$\phi$,$k$)\;
            \ForEach{$\phi'\in S_k'$}{
                $\mathit{flag}\gets$ check\_acceptable($\phi'$,$\Omega_+$)\;
                \If{$\mathit{flag}=\top$}{
                    $S_+$.add($\phi'$)\;
                }
            }
        }
        \Return $S_+$\;
\end{algorithm}

\noindent\textbf{Trivial enumeration.} The basic idea is that we enumerate $p$TRE with increasing length from $k=1$ and check whether the current $p$TRE is a $p$TRE$_+$. More precisely, we start from a placeholder $\hole$ and only utilize the last three rules to generate all $p$TRE with a given length $k$. Therefore, at this stage, each $p$TRE contains no concrete actions. We write $S_k$ for the set of such $p$TRE. For each $k$-length $p$TRE in $S_k$, we utilize the first rule to substitute all placeholders and thus get a finite set of closed $p$TRE $S'_k$. Subsequently, we can select out a $k$-length $p$TRE$_+$ by checking whether a closed $p$TRE accepts all positive examples. Algorithm~\ref{algorithm: trivial} summarizes trivial enumeration for $p$TRE$_+$ with length $k\in\mathbb{N}$. We have the following conclusion. 

\begin{myexample}
Suppose the alphabet $\Sigma=\{a,b\}$. For $k=2$, we first use Rule 2 $\hole\rightarrow \tr{\hole^*}_{\mathbb{I}}$ to get $S_2=\{\tr{\hole^*}_{\mathbb{I}}\}$. For $k=3$, we utilize the last three rules to get $S_3=\{\tr{\tr{\hole^*}_{\mathbb{I}}^*}_{\mathbb{I}},\tr{\hole\cdot\hole}_{\mathbb{I}}, \tr{\hole\vee\hole}_{\mathbb{I}}\}$. Then we utilize the first rule to get $S'_2=\{\tr{\tr{\underline{a}}_{\mathbb{I}}^*}_{\mathbb{I}}, \tr{\tr{\underline{b}}_{\mathbb{I}}^*}_{\mathbb{I}}\}$ and $S_3=\{\tr{\tr{\tr{\underline{a}}_{\mathbb{I}}^*}_{\mathbb{I}}^*}_{\mathbb{I}},\tr{\tr{\tr{\underline{b}}_{\mathbb{I}}^*}_{\mathbb{I}}^*}_{\mathbb{I}},\tr{\tr{\underline{a}}_{\mathbb{I}}\cdot\tr{\underline{a}}_{\mathbb{I}}}_{\mathbb{I}},\tr{\tr{\underline{a}}_{\mathbb{I}}\cdot\tr{\underline{b}}_{\mathbb{I}}}_{\mathbb{I}}$, $\tr{\tr{\underline{b}}_{\mathbb{I}}\cdot\tr{\underline{a}}_{\mathbb{I}}}_{\mathbb{I}},\tr{\tr{\underline{b}}_{\mathbb{I}}\cdot\tr{\underline{b}}_{\mathbb{I}}}_{\mathbb{I}},\tr{\tr{\underline{a}}_{\mathbb{I}}\vee\tr{\underline{a}}_{\mathbb{I}}}_{\mathbb{I}},\\\tr{\tr{\underline{a}}_{\mathbb{I}}\vee\tr{\underline{b}}_{\mathbb{I}}}_{\mathbb{I}}$, $\tr{\tr{\underline{b}}_{\mathbb{I}}\vee\tr{\underline{a}}_{\mathbb{I}}}_{\mathbb{I}},\tr{\tr{\underline{b}}_{\mathbb{I}}\vee\tr{\underline{b}}_{\mathbb{I}}}_{\mathbb{I}}\}$.
\end{myexample}

\begin{mylemma}\label{lemma:trivial_gen}
All $p$TRE$_+$ with length $k\in\mathbb{N}$ w.r.t a given positive examples $\Omega_+$ can be generated by Algorithm~\ref{algorithm: trivial}.
\end{mylemma}

\noindent\textbf{Recursive Generation with Edge Pruning.} The intuition of trivial enumeration is that all $p$TREs are generated and checked one-by-one to be $p$TRE$_+$. Here we propose a method based on recursive generation and edge pruning. Instead of collecting all $p$TRE with length $k$ first, we recursively generate a closed $p$TRE with length $k$ based on all substitution rules, and prune the edge $p$TRE which is impossible to expand to $p$TRE$_+$.



\begin{algorithm}[t]\label{algorithm: recursive}
	\caption{Recursive Enumeration with Edge Pruning}
	\label{alg:encoding_path}
	\SetKwInOut{Input}{input}
	\SetKwInOut{Output}{output}
	\Input{A required length $k$; the positive samples $\Omega_+$}
	\Output{The set of $k$-length $p$TRE$_+$ $S_+$.}
	$\mathit{S_0}\gets \mathit \hole$; $\mathit{step}\gets 0$\; 
	\While{$\mathit{step} < k$}{
	    \ForEach{ $\mathit{\phi}\in\mathit{S_0}$}{
            \If{$\mathit{\phi}$ is not closed}{
            $\mathit{S_1}$.add($\mathit{\phi}$'s direct children)\;
            }}
        $\mathit{step}\gets{\mathit{step}+1}$\;
        
        \ForEach{ $\mathit{\phi}\in\mathit{S_1}$}{
        \If{$\phi$ is an edge $p$TRE}{
            edge\_pruning($\mathit{\phi}$)\;
            }
        }
        $\mathit{S_0} \gets \mathit {S_1}$; $\mathit{S_1}\gets \emptyset$\;
        
	}
    \ForEach{ $\mathit{\phi}\in\mathit{S_0}$}{
    \If{$\mathit{\phi}$ is closed}{
           $\mathit{flag}\gets$ check\_acceptable($\phi$,$\Omega_+$)\;
                \If{$\mathit{flag}=\top$}{
                    $S_+$.add($\phi$)\;
                }
            }
    }
	\Return $\mathit{S}_+$\;
\end{algorithm}

Here we call $p$TRE $\phi_1$ to be one of $p$TRE $\phi_2$'s (direct) children iff $\phi_1$ can be generated from $\phi_2$ by arbitrarily applying generating rules for (one) times. The $p$TREs are not enumerated according to length as it is in trivial enumeration, but in the number of recursive steps from the start $\phi=\hole$. 
Once reaching a $k$-length closed $p$TRE, we check whether it is a $p$TRE$_+$. 

During the recursive generation, if the current $p$TRE is an edge $p$TRE (see the definition in Section~\ref{sbsc:overview}), instead of utilizing Rule $\hole\rightarrow \tr{\underline{\sigma}}_{\mathbb{I}}$ directly, we replace the only placeholder with a special formula $\tr{\Sigma^*}_{[0,\infty)}$, and thus the resulting closed $p$TRE can be viewed as an over-approximation of all possible children of the current edge $p$TRE. This adapts the idea in~\cite{LeeSO16} to our TRE settings. If this over-approximation fails to accepted all positive examples, we can conclude that all children of the edge $p$TRE are impossible to be $p$TRE$_+$, and therefore we prune the edge $p$TRE directly. The whole process is presented in Algorithm~\ref{algorithm: recursive}. We have the following results.

\begin{mytheorem}\label{thm:recursive_gen_len}
A $p$TRE$_+$ generated after $k$ steps is of length $k$. All $k$-length $p$TRE$_+$ are generated at the $k$-th step in the recursive generation.
\end{mytheorem}

\begin{myexample}
Consider the set of positive examples $\{\omega_1=(a,1.2)(a,2.2)$, $\omega_2=(b,1.2)(a,2.6)\}$. The edge $p$TRE $\tr{\underline{a}_{\mathbb{I}}\cdot \hole}_{\mathbb{I}}$ can be pruned before substituting the last placeholder, since $(\underline{a}_{\mathbb{I}}\cdot \tr{\Sigma^*}_{[0,\infty)})_{\mathbb{I}}$ doesn't accept $\omega_2$. $((\underline{a}_{\mathbb{I}}\vee\underline{b}_{\mathbb{I}})_{\mathbb{I}}\cdot \hole)_{\mathbb{I}}$ can not be pruned using the above strategy, since $\tr{\tr{\underline{a}_{\mathbb{I}}\vee\underline{b}_{\mathbb{I}})_{\mathbb{I}}\cdot \tr{\Sigma^*}_{[0,\infty)}}}_{\mathbb{I}}$ accept both $\omega_1$ and $\omega_2$. 
\end{myexample}


\noindent\textbf{Containment Pruning.}
The recursive-style enumeration is more friendly to pruning. However, it brings a disadvantage of repeatedly generating $p$TREs. For instance, $((\underline{a}_{\mathbb{I}}\cdot \hole)_{\mathbb{I}} \cdot (\hole \cdot \underline{a}_{\mathbb{I}})_{\mathbb{I}})_{\mathbb{I}}$ can be generated directly from $((\underline{a}_{\mathbb{I}}\cdot \hole)_{\mathbb{I}} \cdot (\hole \cdot \hole)_{\mathbb{I}})_{\mathbb{I}}$ and $((\hole \cdot \hole)_{\mathbb{I}} \cdot (\hole \cdot \underline{a}_{\mathbb{I}})_{\mathbb{I}})_{\mathbb{I}}$.
We introduce a strategy, named containment pruning, of which the insight is to remember the pruned $p$TREs by the edge pruning and to prevent generating their children from other $p$TREs. The pruned $p$TREs are stored in a set $S_{\mathit{doomed}}$. When a new $p$TRE is generated, it will be checked whether it is contained by a $p$TRE in $S_{\mathit{doomed}}$. The containment can be decided utilizing the methods proposed in~\cite{HUNT1976222,doi:10.1137/0214044,10.1145/1925844.1926429}. If so, the new $p$TRE is pruned from the recursive generation and added to $S_{\mathit{doomed}}$. The whole process is presented in Algorithm~\ref{algorithm: containment}.

\begin{algorithm}\label{algorithm: containment}
	\caption{Recursive Enumeration with Edge Pruning and Containment Pruning}
	\label{alg:containment}
	\SetKwInOut{Input}{input}
	\SetKwInOut{Output}{output}
	\Input{required length $k$; alphabet $\Sigma$}
	\Output{The set of all $k$-length $p$TRE$_+$ $S_+$.}
	$\mathit{S_0}\gets \mathit \hole$; $\mathit{step}\gets 0$\; 
	\While{$\mathit{step} < k$}{
	    \For{ $\mathit{\phi}\in\mathit{S_0}$}{
            \If{$\mathit{\phi}$ is not closed}{
            $\mathit{S_1}$.add$(\mathit{\phi}'s$ direct children)\;
            }}
        $\mathit{step}\gets{\mathit{step}+1}$\;
        
        \For{$\mathit{\phi}\in\mathit{S_1}$}{
        \If{$\phi$ is an edge $p$TRE}{
            \textbf{do} edge pruning on $\mathit{\phi}$\;

            \If{$\mathit{\phi}$ is pruned}{$\mathit{S_{doomed}}$.add$(\mathit{\phi}$)\;}
            }

        \textbf{do} RE containment pruning on $\mathit{p}$\;
            \If{$\mathit{\phi}$ is pruned}{$\mathit{S_{doomed}}$.add$(\mathit{\phi}$)\;}
            }
        $\mathit{S_0}\gets \mathit {S_1}$\;$\mathit{S_1}\gets \{\}$\;
        
	}
    \ForEach{ $\mathit{\phi}\in\mathit{S_0}$}{
    \If{$\mathit{\phi}$ is closed}{
           $\mathit{flag}\gets$ check\_acceptable($\phi$,$\Omega_+$)\;
                \If{$\mathit{flag}=\top$}{
                    $S_+$.add($\phi$)\;
                }
            }
    }
	\Return $\mathit{S}$
\end{algorithm}
\begin{myremark}
The containment problem is of PSPACE-complete~\cite{HUNT1976222}. Hence, it is an expensive strategy. When and how frequently it should be utilized still needs to be explored. 
\end{myremark}


\subsection{Encoding and solving time restrictions}\label{sc:time_cons}

Suppose that we have a $p$TRE$_+$ $\phi$ that accepts all possible examples. In this section, we present an SMT-based method to filter all negative examples by synthesizing a set of instances of the parametric intervals in $\phi$. For a negative example $\omega_-\in\Omega_-$, there are two situations w.r.t $\phi$ as follows. If $\omega_-\not\in \semantic{\phi}$, we do nothing. If $\omega_-\in\semantic{\phi}$, it means that the untimed sequence $\mu(\omega_-)$ matches the corresponding RE $\mu(\phi)$, and we need to find a set of instances so that the resulting TRE $\varphi$ can reject it. We also need to keep $\varphi$ accepting all positive examples with these instances. If we can not find such a satisfiable TRE $\varphi$, we drop this $\phi$ and try the next candidate $p$TRE$_+$.


Let us start with an illustrative example for a negative example $\omega_-\in\Omega_-$ such that $\omega_-\in\semantic{\phi}$. 

\begin{myexample} \label{example:negative_parsing}
Consider a negative example $\omega_- = (a,1.5)(b,2)\\(b,3)$ and a $p$TRE$_+$
$\phi=\tr{\tr{\tr{\underline{a}}_{\mathbb{I}_4} \vee \tr{\tr{\underline{a}}_{\mathbb{I}_7}\cdot\tr{\underline{b}}_{\mathbb{I}_8}}_{\mathbb{I}_5}}_{\mathbb{I}_{2}} \cdot \tr{\tr{\underline{b}}_{\mathbb{I}_6}^*}_{\mathbb{I}_3}}_{\mathbb{I}_1}$, we have the untimed version $\mu(\omega_-) = abb$ and $\mu(\phi) = (a \vee ab)\cdot b^*$. It is obvious that $\mu(\omega_-)$ matches $\mu(\phi)$ with two parsing solutions. One is that $\omega$ is composed with the symbol `$a$' at left of $\vee$, then `$b^*$' repeats for 2 times. The other is `$ab$' followed by one iteration of `$b^*$'. Therefore, if we need the resulting TRE $\varphi$ to reject $\omega_-$, the instances of parametric intervals should reject both parsing solutions by filtering the timing information in $\omega_-$. 
\end{myexample}

Hence, we need to find all parsing solutions of $\mu(\omega_-)$ w.r.t the untimed RE $\mu(\phi)$. For our particular encoding problem, the basic idea is as follows. 
\begin{enumerate}
    \item Label each symbol in $\mu(\phi)$ with a subscript which is the position of the corresponding node on the syntax tree, thus obtaining a linear expression $\psi$.
    \item Transform $\psi$ to an nondeterministic finite automata (NFA) utilizing Glushkov's construction \cite{Glu1961Abstract}, which is to construct an NFA from a RE. In a Glushkov's NFA, each transition action is labeled by a subscripted symbol of a leaf node of the syntax tree. Each \emph{accepting path} (defined below) of the NFA, by taking the labels of all transitions of an into a sequence, represents how the symbols of an untimed word are matched to the leaf nodes of the tree. 
    \item Find all parsing solutions w.r.t $\mu(\omega_-)$ and $\mu(\phi)$, each is represented as an accepting path. 
    \item For each path $p$, we encode the requirement that the time information in $\omega_-$ does not satisfy the time restrictions along the path into an SMT formula $\gamma$. 
\end{enumerate}

An accepting path $p$ of the Glushkov's NFA for $\phi$ is a finite sequence of pairs $(x,d)$, where $x$ is the subscript of a symbol in $\phi$ and $d$ is the depth of the corresponding node on the syntax tree. We write $p[i]$ for the element at the $i$-th index of $p$.

\begin{figure}[t]
    \begin{minipage}{0.45\linewidth}
    \centering
    \resizebox{\linewidth}{!}{
    \begin{tikzpicture}[sibling distance=2cm, level distance = 1cm, semithick]
    \node (1) {\Large $\cdot$}
        child {node (2) {$\vee$}
            child {node (4) {$a$}}
            child {node (5) {\Large $\cdot$}
                child {node (7) {$a$}}
                child {node (8) {$b$}}
            }
        }
        child {node (3) {\Large $^*$} 
            child {node (6) {$b$}}
        };
    \node[right = .2cm, red!50] at (1) {1};
    \node[right = .2cm, red!50] at (2) {2};
    \node[right = .2cm, red!50] at (3) {3};
    \node[right = .2cm, red!50] at (4) {4};
    \node[right = .2cm, red!50] at (5) {5};
    \node[right = .2cm, red!50] at (6) {6};
    \node[right = .2cm, red!50] at (7) {7};
    \node[right = .2cm, red!50] at (8) {8};
    \end{tikzpicture}
    }
    \begin{center}
    \vspace{-0.2cm}
    (a) 
    \vspace*{0.2cm}
    \end{center}
    \end{minipage}
    \begin{minipage}{0.48\linewidth}
    \centering
    \resizebox{\textwidth}{!}{
    \begin{tikzpicture}[->, >=stealth', shorten >=1pt, auto, node distance=1.5cm, semithick,scale=0.8, every node/.style={scale=0.8}, initial where=above]
    \node[initial,state] (0) {\large $q_0$};
    \node[accepting, state] (4) [right = 1.2cm of 0] {\large $q_4$};
    \node[state] (7) [left = 1.2cm of 0] {\large $q_7$};
    \node[accepting, state] (6) [below = 1.2cm of 0] {\large $q_6$};
    \node[accepting, state] (8) [below = 1.2cm of 7] {\large $q_8$};
    \path (0) edge node[above] {\large $a_7$} (7)
    (0) edge node[above] {\large $a_4$} (4)
    (0) edge node[right] {\large $b_6$} (6)
    (7) edge node[right] {\large $b_8$} (8)
    (8) edge node[above] {\large $b_6$} (6)
    (4) edge node[right=2pt] {\large $b_6$} (6)
    (6) edge[in= -40, out=0, loop] node[right] {\large $b_6$} (6);
    \end{tikzpicture}
    }
    \begin{center}
    \vspace{-0.1cm}
    (b) 
    \vspace*{0.2cm}
    \end{center}
    \end{minipage}
    \begin{minipage}{\linewidth}
    \centering
    \resizebox{0.6\linewidth}{!}{
    \begin{tikzpicture}[<-, sibling distance=4cm, level distance = 2cm, semithick]
    \node[align=center] (1) {\Large $1$ \\ \large ($6.5\in\mathbb{I}_{1}$)} [grow=up]
        child {node[align=center] (2) {\large $3$ \\ \large ($3\in\mathbb{I}_{3}$)}
            child {node[align=center] (4) {\large $6$ \\ \large ($3\in\mathbb{I}_6$)}
            }
        }
        child {node[align=center] (3) {\large $2$ \\ \large ($3.5\in\mathbb{I}_{2}$)}
            child {node[align=center] (5) {\large $5$ \\ \large ($3.5\in\mathbb{I}_{5}$)}
                child {node[align=center] (7) {\large $8$ \\ \large ($2\in\mathbb{I}_{8}$)}}
                child {node[align=center] (8) {\large $7$ \\ \large ($1.5\in\mathbb{I}_{7}$)}}
            }
        };
    \end{tikzpicture}
    }
    \end{minipage}
    \begin{center}
    (c) 
    \end{center}
    \caption{(a): the syntax tree of $\mu(\phi) = (a \vee ab)\cdot b^*$; (b): the corresponding Glushkov's NFA $A$ of $\psi = (a_4 \vee a_7 b_8)\cdot b_6^*$; (c): the encoding process of path $7,8,6$.}
    \label{fig:encoding}
\end{figure}
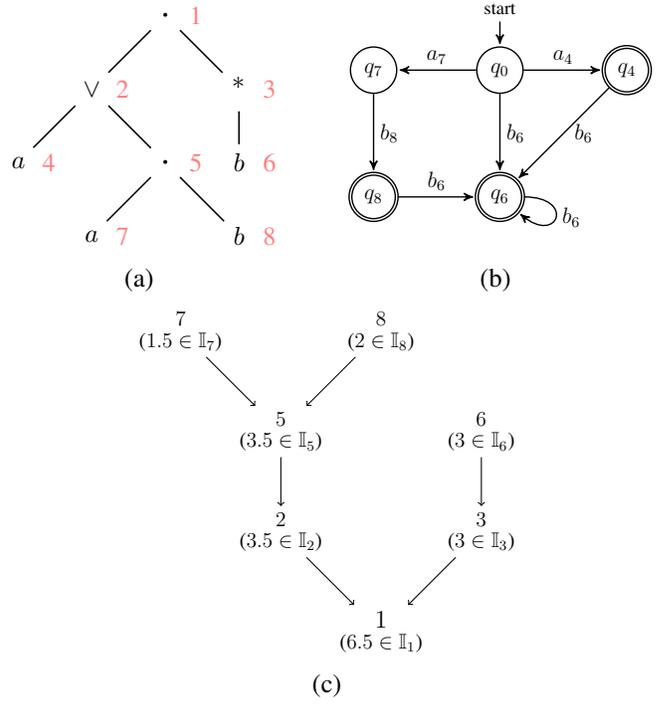

\begin{myexample} \label{example:expression_NFA}
Continue Example~\ref{example:negative_parsing}. Given $\mu(\phi) = (a \vee ab)\cdot b^*$, we have the corresponding syntax tree in Fig.~\ref{fig:encoding}. By renaming the symbols, we get a linear expression $\psi= (a_4 \vee a_7 b_8)\cdot b_6^*$. The subscript is the position of the corresponding symbols on the syntax tree. Then we build a Glushkov's NFA $A$ w.r.t $\psi$ as shown in Fig.~\ref{fig:encoding}. For the string $\mu(\omega_-)=abb$, two accepting paths $(4,3),(6,3),(6,3)$ and $(7,4),(8,4),(6,3)$ are found. For each path, based on the semantics of TRE, we can encode the requirement that the corresponding parametric intervals in $\phi$ filter the path. 
\end{myexample}

\begin{algorithm}[!t]
    \small
	\caption{Encoding path formula}
	\label{alg:encoding_path}
	\SetKwInOut{Input}{input}
	\SetKwInOut{Output}{output}
	\Input{a negative example $\omega_-$; a $p$TRE$_+$ $\phi$; an accepting path $p$.}
    \Output{a path formula $\gamma$.}
    $\mathit{ctimes}\gets \lambda(\omega)$; \label{alg_encoding:ctimes} $\mathit{depth}\gets \mathit{max\_d}(p)$, $C \gets \emptyset$\;
    \For{$i\in [0,|p|-1]$}{
        $\beta \gets $ encode\_single\_node($\mathit{ctimes}$, $p$, $i$); $C$.add($\beta$)\; \label{alg_encoding:encode_single_node}
    }
    \While{$\mathit{depth} \neq 1$}{
	    $i \gets 0$;
        $\mathit{next\_p}\gets[]$; $\mathit{next\_{time}}\gets[]$\;
	    \While{$i < |p|$}{
	        $\mathit{cpos}\gets p[i].x$\; $\mathit{cnode}\gets\phi[\mathit{cpos}]$\; $\mathit{next\_pos}\gets p[i+1].x$\;  $\mathit{next\_node}\gets\phi[\mathit{next\_pos}]$\;
	        \If{$\mathit{cnode}.\mathit{depth} = \mathit{depth}$}{
	            \If{$\mathit{cnode}$.parent is a disjunction node}{
                    $\mathit{next\_p}$.add($\mathit{cnode.parent}$)\; \label{alg_encoding:disjunct_merge}
                    $\mathit{next\_{time}}$.add($\mathit{ctimes}[i]$)\;
                    $i\gets i+1$\;
                    }
	            \If{$\mathit{cnode}$.parent is a concatenation node \textbf{and} 
             $\mathit{next\_node}$.parent $=$ $\mathit{cnode.parent}$}{
                    $\mathit{next\_p}$.add($\mathit{cnode.parent}$)\; \label{alg_encoding:concat_merge}
                    $\mathit{next\_{time}}$.add($\mathit{ctimes}[i]+\mathit{ctimes}[i+1]$)\;
                    $i\gets i+1$\;
                    }
	            \If{$\mathit{cnode}$.parent is a star node}{
                    \textbf{find} the first $j>i$ s.t. $\phi[p[j].x]$.parent$ \neq \mathit{cnode}$.parent \textbf{or} $j=|p|$\;
                    $\mathit{next\_p}$.add($\mathit{cnode.parent}$)\;
                    $\mathit{next\_{time}}$.add($\sum_{i\leq k < j}ctime[k]$)\;
                    $i\gets j$\;
                    }
                    $\beta \gets $encode\_single\_node($\mathit{ctimes}$, $p$, $i$)\; $C$.add($\beta$)\;
	        }
            \Else {$\mathit{next\_p}.add(p[i])$\;$\mathit{next\_{time}}.add(ctime[i])$\;$i\gets i+1$\;}
            
	    }
     $p\gets\mathit{next\_p}$\; $\mathit{ctimes}\gets \mathit{next\_{time}}$\;
     $\mathit{depth}\gets \mathit{max\_d}(p)$\;
    }
    $\gamma \gets \bigwedge_{\beta\in C}\neg\beta$\;
    \Return $\gamma$ \;
\end{algorithm}

Algorithm \ref{alg:encoding_path} depicts the process to encode the requirement that a $p$TRE$_+$ $\phi$ rejects an 
accepting path $p$ w.r.t a negative example $\omega_-$ into an SMT formula $\gamma$. Note that to encode the requirement that a positive example $\omega_+$ is accepted by $\phi$ is similar, except that at the end of the algorithm it is the disjunction of all $\beta$ instead of the conjunction of $\neg \beta$ that is constructed.
Function parent($\phi$, $p$, $i$) returns a pair $(x',d')$, where $x'$ is the subscript of $\phi[p[i].x].\mathit{parent}$ and $d'=p[i].d-1$. And function encode\_single\_node($\mathit{ctimes}$, $p$, $i$) is to generate constraint $\mathit{ctimes}[i]\in \mathbb{I}_{p[i].x}$. At first each symbol of $\omega_-$ is matched to a leaf node on the syntax tree of $\phi$ and the time constraints of these leaf nodes are encoded. Then after each loop the nodes with the max depth in the array $cnode$ are replaced by their parent nodes on the tree, and the time constraints of these parent nodes are encoded. 

\begin{myexample} \label{example:path_encoding}
Continue to consider $\omega_- = (a,1.5)(b,2)(b,3)$, $\phi=\tr{\tr{\tr{\underline{a}}_{\mathbb{I}_4} \vee \tr{\tr{\underline{a}}_{\mathbb{I}_7}\cdot\tr{\underline{b}}_{\mathbb{I}_8}}_{\mathbb{I}_5}}_{\mathbb{I}_{2}} \cdot \tr{\tr{\underline{b}}_{\mathbb{I}_6}^*}_{\mathbb{I}_3}}_{\mathbb{I}_1}$, and one accepting path $p = (7,4),(8,4),(6,3)$. We have $\mathit{ctimes} = 1.5,2,3$ at the beginning. After encoding the timing constraints for leaf nodes at the positions $7,8,6$ (Line~\ref{alg_encoding:encode_single_node}), we obtain the formulas $1.5\in\mathbb{I}_{7}$, $2\in\mathbb{I}_{8}$, and $3\in\mathbb{I}_{6}$. Since the father node of the two nodes at position $7$ and $8$ is the concatenation node at position $5$, we have $\mathit{next\_p}=(5,3),(6,3)$ and $\mathit{next\_ctimes} = 3.5,3$, and thereby obtain a new formula $3.5\in\mathbb{I}_5$ (Line~\ref{alg_encoding:concat_merge}). Since the father node of the node at position $5$ is the disjunction node at position $2$, we have $\mathit{next\_p}=(2,2),(6,3)$ and $\mathit{next\_ctimes} = [3.5,3]$, and obtain a new formula $3.5\in\mathbb{I}_2$  (Line~\ref{alg_encoding:disjunct_merge}). With the same process, we obtain the formulas $3\in\mathbb{I}_3$ and $6.5\in\mathbb{I}_1$ for the time restrictions of the nodes at positions $3$ and $1$. Finally, we get a path formula $\gamma = \neg(1.5\in\mathbb{I}_{7} \wedge 2\in\mathbb{I}_{8} \wedge 3\in\mathbb{I}_{6} \wedge 3.5\in\mathbb{I}_5 \wedge 3.5\in\mathbb{I}_2 \wedge 3\in\mathbb{I}_3 \wedge 6.5\in\mathbb{I}_1)$. 
\end{myexample}

In general, the basic encoding process of the SMT problem $\Phi_{\Omega}$ with $p$TRE$_+$ $\phi$ and examples $\Omega$ is as follows:
\begin{itemize}
    \item Positive examples encoding: For each positive example $\omega_+$, we first solve the parsing problem whether $\omega_+\in\semantic{\phi}$, and find all accepting paths. Once the paths are identified, encode the time constraints associated with each path into formulas. Subsequently, construct the disjunction of path formulas to form the constraints for a single positive example.
    \item Negative examples encoding: For each negative example $\omega_-$, we first check whether it is consistent with $p$TRE$_+$ $\phi$. If not, then we do not need to parse it. If it is consistent, proceed to solve the parsing problem $\omega_-\in\semantic{\phi}?$ and find all accepting paths. Once the paths are identified, encode the time constraints associated with each path into formulas. Subsequently, construct the conjunction of the negations of all path formulas to form the constraints for a single negative example.
    \item SMT problem construction: $\Phi_{\Omega}$ is the conjunction of all constraints encoded from all positive and negative examples.
\end{itemize}

\begin{mytheorem}\label{thm:ptre_formula_sat}
Given a $p$TRE$_+$ $\phi$, there exists a TRE $\varphi\in[\phi]$ consistent with $\Omega$ if and only if $\Phi_{\Omega}$ is satisfiable.
\end{mytheorem}

\section{Experiments and Implementation.} \label{sc:experiments}
We have implemented our synthesis methods proposed in Section \ref{sc:syn_pro}, including the strategies for enumerating and pruning $p$TRE as proposed in Section \ref{sc:enum_and_prune} along with the timed constraints encoding method detailed in Section \ref{sc:time_cons}. The prototype is developed in Python 3.8. We use Z3~\cite{MouraB08} as our SMT solver. All experiments were conducted on a laptop:
\begin{itemize}
    \item Processor: 11th Gen Intel(R) Core(TM) i5-11300H @ 3.10GHz, with 4 cores and 8 logical processors.
    \item RAM: 16GB installed
\end{itemize}

Note that the correctness of our synthesis method is guaranteed by the proposed theorems. 
Therefore, we conducted three kinds of experiments to demonstrate its effectiveness and efficiency. First, we compared the efficiency of the three synthesis strategies. Second, we demonstrated the scalability of our synthesis process on randomly generated timed words. 
At last, we evaluated our synthesis method using a case study involving a train gate system~\cite{10.1007/978-3-031-06773-0_26} modeled by a timed automaton. 
We utilized the timed automata uniform sampler $\bf{WORDGEN}$~\cite{10.1145/3575870.3587116} to generate timed words. 

\subsection{The Comparison of Strategies}
\begin{figure}
    \centering
    \resizebox{0.75\linewidth}{!}{
    \begin{tikzpicture}[->, >=stealth', shorten >=1pt, auto, node distance=5cm, semithick,scale=0.6, every node/.style={scale=0.8}, initial where=above]
    \node[initial,state] (0) {\large $q_0$};
    \node[state] (1) [below left = 2.5cm of 0] {\large $q_1$};
    \node[accepting, state] (2) [below right = 2.5cm of 0] {\large $q_2$};
    \path (0) edge [loop left] node {\large $a,[0,2)$} (0)
    (0) edge[in = 110, out = -20] node {\large $a,[2,4)$} (2)
    (0) edge node[below right] {\large $b,[0,4)$} (1)
    (1) edge [in = -160, out = 70] node {\large $b,[0,4)$} (0)
    (1) edge [in= -160, out=-20] node[below] {\large $a,[0,4)$} (2)
    (2) edge [left] node {\large $a,[2,4)$} (0)
    (2) edge [in= 0, out=180] node[above] {\large $a,[0,2)$} (1)
    (2) edge[in= -40, out=0, loop] node[right] {\large $b,[0,4)$} (2);
    \end{tikzpicture}
    }
    \caption{The timed automaton used to mark positive and negative examples in the strategies comparison.}
    \label{fig:TA1}
\end{figure}
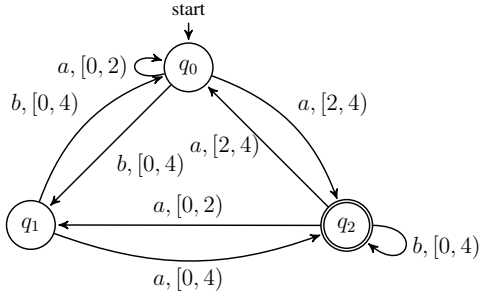

In this subsection, we evaluate and compare the different strategies: recursive with edge pruning and recursive with containment pruning. The experiments were conducted using a setting of alphabet $\Sigma = \{a,b\}$ and time delay $t\in [0,4)$. Timed words of selected lengths are randomly generated and fed into a one-clock and full-reset timed automaton (as depicted in Fig. \ref{fig:TA1}). Accepted timed words are marked as positive examples, while rejected ones are marked as negative. This process is repeated until we get the desired \emph{number of positive examples (nop)}. 
In the experiment, around $40\%$ of the randomly generated timed words are marked positive. 
As the randomly generated timed words are not enough to characterize the timed automaton, we don't require that the synthesized TRE exactly matches the corresponding TRE of the automaton.


Strategy setting: 
\begin{itemize}
    \item Edge pruning strategy: This strategy involves pruning only on edge pTREs.
    \item Containment Pruning Strategy: In this strategy, we perform a partial containment check. Specifically, we verify if the syntax tree of a longer pTRE contains that of a shorter one, excluding placeholders $\hole$.
\end{itemize}

\begin{table}
\caption{Experimental results for different strategies. $\textbf{Above}$: trivial. $\textbf{Middle}$: recursive $\&$ edge pruning. $\textbf{Below}$: recursive $\&$ containment pruning.}
\label{table:experiment_results_strategies}
    \centering
    \begin{tabular}{|c|c|c|c|c|c|c|}
    \hline
    \diagbox{length}{time(s)}{nop}   & 5 & 7 & 9 & 11 & 13 & 15 \\ \hline
6  &  0.61& 146.3  & 0.38  &  295.6  &  173.5  &  167.8  \\ \hline
7  &  47.3 &  169.6 & 169.4  &  161.4  &  170.0  &  172.2  \\ \hline
8  & 1.75  &  48.9 &  194.1 &  165.9  &   173.6 &  171.3  \\ \hline
9  &  160.5 &  1.76 &  172.1 &   169.6 &  170.1  &  167.5  \\ \hline
10 & 156.1  &  169.7 &  175.8 &  172.4  &   160.3 &  176.7  \\ \hline
\end{tabular}

\vspace{0.5cm}

    \begin{tabular}{|c|c|c|c|c|c|c|}
    \hline
\diagbox{length}{time(s)}{nop}   & 5 & 7 & 9 & 11 & 13 & 15 \\ \hline
6  &  0.46& 114.7  & 0.35  &  236.5  &  130.7  &  128.4  \\ \hline
7  &  36.2 &  130.4 & 130.8  &  122.9  &  123.9  &  129.2  \\ \hline
8  & 1.55  &  35.4 &  153.2 &  125.3  &   127.6 &  127.8  \\ \hline
9  &  126.8 &  1.50 &  129.0 &   127.6 &  124.1  &  119.5  \\ \hline
10 & 117.2  &  132.3 &  133.1 &  132.3  &   119.4 &  124.0  \\ \hline
\end{tabular}

\vspace{0.5cm}
    \begin{tabular}{|c|c|c|c|c|c|c|}
    \hline
\diagbox{length}{time(s)}{nop}   & 5 & 7 & 9 & 11 & 13 & 15 \\ \hline
6  &  0.41& 107.8  & 0.33  &  245.9  &  125.8  &  122.5  \\ \hline
7  &  32.6 &  128.5 & 125.7  &  120.1  &  115.0  &  120.6  \\ \hline
8  & 1.52  &  32.1 &  154.7 &  119.2  &   124.9 &  118.8  \\ \hline
9  &  122.5 &  1.48 &  124.3 &   122.4 &  117.5  &  120.1  \\ \hline
10 & 112.7  &  127.0 &  127.2 &  124.8  &   116.4 &  123.2  \\ \hline
\end{tabular}
\end{table}

Table~\ref{table:experiment_results_strategies} presents the experimental results by using different strategies. 
As the length of the examples increases, longer examples tend to produce more complex TRE with minimal length. Consequently, the number of $p$TREs requiring enumeration and verification grows exponentially, leading to a significant increase in computational time. Additionally, the results indicate that pruning strategies generally demonstrate superior performance compared to the trivial strategy. Nonetheless, the implementation of more sophisticated pruning techniques does not necessarily result in enhanced search efficiency. This may be attributed to the increased time required for containment checks as the set of identified $p$TREs expands.

\subsection{Scalability}
In this subsection, we evaluate the scalability of our method. 
We prepare 16 TREs of which the length $L$ is scaled from 6 to 9 and the size of alphabet $|\Sigma|$ is scaled from 2 to 5, as listed in Table \ref{tab:TREs_1}. For conciseness and readability, the time intervals in TREs are abbreviated as “$\mathbb{I}$”, and brackets $"\tr{}"$ containing a single letter are omitted in the table. In experiment, each “$\mathbb{I}$” is randomly assigned with $(a,b]$, where $0\leq a<b\leq 10, a,b\in\mathbb{N}$, to form valid TREs. 
Then we transform these TREs into corresponding TAs. Positive examples are sampled directly from these TAs, and negative ones are sampled from complemented TAs. The maximum of the upper bounds of guards in complemented TAs is 15. We use $\bf{WORDGEN}$ to generate examples. The range of example lengths is $l\leq8,11,14$ (some of the TREs accept only timed words shorter than 8, so we only test them in group $l\leq8$). For each TRE with a different length of examples, we conducted 3 trials. And for each trial, 300 positive and 300 negative examples were sampled. The time limit of each trial is 1800 seconds. 

\begin{table}[t]
\caption{TREs Used in Scalability Experiment}
\begin{center}
\begin{tabular}{|c|p{2.8cm}<{\centering}|p{2.8cm}<{\centering}|}
\hline
\textbf{Length}&\multicolumn{2}{c|}{\textbf{Size of Alphabet $|\Sigma|$}} \\
\cline{2-3} 
\textbf{of TRE} & 2 & 3  \\
\hline
6& ${\tr{\underline{a}_\mathbb{I}\cdot \tr{\underline{b}_\mathbb{I}}^*}_\mathbb{I}\cdot \underline{a}_\mathbb{I}}$& $\tr{\underline{a}\vee \underline{b}}_\mathbb{I} \cdot \tr{\underline{c}^*}_\mathbb{I}$\\
\hline
7& $\tr{\underline{a}_\mathbb{I}\cdot \underline{b}_\mathbb{I}}\vee \tr{\underline{b}_\mathbb{I}\cdot \underline{a}_\mathbb{I}}$& ${\tr{\underline{a}\vee \underline{b}}}^*_\mathbb{I}\cdot \underline{c}_\mathbb{I}^*$\\
\hline
8&$\tr{\underline{a}_\mathbb{I}\cdot \underline{b}_\mathbb{I}}^*\cdot \tr{\underline{a}\vee \underline{b}}_\mathbb{I}$& $\tr{\underline{a}_\mathbb{I}\cdot \underline{b}_\mathbb{I}\cdot \underline{c}_\mathbb{I}}^*\cdot \underline{a}_\mathbb{I}$\\
\hline
9&$\tr{\underline{a}^*}_\mathbb{I}\cdot \tr{\underline{b}^*}_\mathbb{I}\cdot\tr{\underline{a}\cdot \underline{b}}_\mathbb{I}$&$\tr{\underline{a}\cdot \underline{b}}_\mathbb{I}\cdot\tr{\tr{\underline{c}\cdot \underline{b}}_\mathbb{I}\vee \underline{b}_\mathbb{I}}$ \\
\hline
\end{tabular}
\vspace{0.5cm}

\begin{tabular}{|c|p{2.8cm}<{\centering}|p{2.8cm}<{\centering}|}
\hline
\textbf{Length}&\multicolumn{2}{c|}{\textbf{Size of Alphabet $|\Sigma|$}} \\
\cline{2-3} 
\textbf{of TRE} &  4 &5 \\
\hline
6& $\tr{\tr{\underline{a}^*\vee \underline{b}}_\mathbb{I}\cdot \underline{c}}_\mathbb{I}$&${{\tr{\underline{a}_\mathbb{I}}^*}_\mathbb{I}\cdot \underline{b}_\mathbb{I}\vee \underline{c}_\mathbb{I}}_\mathbb{I}$\\
\hline
7& $\tr{\underline{a}_\mathbb{I}\cdot \underline{b}_\mathbb{I}}_\mathbb{I}\vee \tr{\underline{c}_\mathbb{I}\cdot \underline{a}_\mathbb{I}}$&$\tr{{\underline{a}_\mathbb{I}\vee \underline{b}_\mathbb{I}}\cdot \underline{c}}_\mathbb{I}\cdot \underline{a}_\mathbb{I}$\\
\hline
8& $\underline{a}_\mathbb{I}\cdot \underline{b}_\mathbb{I}\cdot\tr{\underline{c}\vee \underline{d}}_\mathbb{I}^*$&$\tr{\tr{\underline{a}\vee \underline{b}}_\mathbb{I}^*}_\mathbb{I}\cdot \tr{\underline{c}\vee \underline{d}}_\mathbb{I}$\\
\hline
9& $\tr{\underline{a}\vee \underline{b}}_\mathbb{I}^*\cdot\tr{c\vee \underline{d}}_\mathbb{I}^*$&$\tr{\underline{a}_\mathbb{I}\cdot \underline{b}\cdot \underline{c}}_\mathbb{I}\cdot \tr{\underline{d}_\mathbb{I}\cdot \underline{e}}_\mathbb{I}$\\
\hline
\end{tabular}
\label{tab:TREs_1}
\end{center}
\end{table}

\begin{table}[tb]
\caption{Average Time Expanse (seconds) in Scalability Experiment. $\textbf{Above}$: length of examples $l\le8$. $\textbf{Middle}$:$l\le11$. $\textbf{Below}$: $l\le14$.}
\label{tab:scal}
\begin{center}
\begin{tabular}{|c|c|c|c|c|}
\hline
\textbf{Length}&\multicolumn{4}{c|}{\textbf{Size of Alphabet $|\Sigma|$}} \\
\cline{2-5} 
\textbf{of TRE} & 2 & 3 &4 &5 \\
\hline
6& 3.8 & 6.2 &17.3 &34.1\\
\hline
7& 26.6 & 37.4 &60.1 &95.2\\
\hline
8& 173.0 & 316.5 &502.7 &772.8\\
\hline
9& 302.5 & 513.6 & 871.8 & 1365\\
\hline
\end{tabular}

\vspace{0.5cm}

\begin{tabular}{|c|c|c|c|c|}
\hline
\textbf{Length}&\multicolumn{4}{c|}{\textbf{Size of Alphabet $|\Sigma|$}} \\
\cline{2-5} 
\textbf{of TRE} & 2 & 3 &4 &5 \\
\hline
6& 4.2 & 7.7 & 22.4 & 46.7\\
\hline
7& - & 46.2 &- &-\\
\hline
8& 198.6 & 342.7 &374.7$^a$ &406.9$^a$\\
\hline
9& 363.8 & - &591.3$^a$ &-\\
\hline
\end{tabular}

\vspace{0.5cm}

\begin{tabular}{|c|c|c|c|c|}
\hline
\textbf{Length}&\multicolumn{4}{c|}{\textbf{Size of Alphabet $|\Sigma|$}} \\
\cline{2-5} 
\textbf{of TRE} & 2 & 3 &4 &5 \\
\hline
6& 4.8 & 9.7 & 27.2 & 63.0\\
\hline
7& - & 58.5 &- &-\\
\hline
8& 236.5 & 170.6$^a$ &419.3$^a$ &157.5$^a$\\
\hline
9& 392.5 & - &662.8$^a$ &-\\
\hline
\end{tabular}
\end{center}
$^{\mathrm{a}}$ Not all 3 trials successfully generate the exact target TRE.
\end{table}

As shown in Table \ref{tab:scal}, our method is capable of synthesizing a TRE in a reasonable time. Along with a larger alphabet, a more complex target TRE, and longer examples, the more time it costs. The size of the alphabet and the length of the target TRE influence the running time hugely since they increase the number of $p$TREs to be enumerated. Length of examples exerts a slight influence on the duration of the synthesis process. 

Note that the synthesis results may not be the same as the target TREs. Sometimes only the time intervals are different, but it is more often that a synthesized TRE is shorter than the corresponding target TRE. This phenomenon can be attributed to the synthesis method always synthesizing the minimal consistent TRE $\varphi$. If there exists a TRE $\varphi$ containing the language of the target TRE $\phi$, i.e., $\semantic{\varphi}\supset \semantic{\phi}$, our method will find such $\varphi$. The shorter TRE $\varphi$ is discarded if and only if the set of negative examples contains a member that is recognized by the shorter TRE but not by the target TRE, i.e. $\exists\omega_-\in\Omega_-, \omega_-\in\semantic{\varphi}\backslash\semantic{\phi}$. The likelihood of encountering a longer negative example or a negative example sampled with larger alphabet such that $\omega_-\in\semantic{\varphi}\backslash\semantic{\phi}$ is smaller. Consequently, when the  number of examples is fixed, it is less probable that the target TRE $\phi$ will be synthesized as the length of examples and size of alphabet increase.

\begin{table}[tb]
\caption{Average Time Expanse (seconds) and  Correct Times in Input Experiment. }
\begin{center}
\begin{tabular}{|c|c|c|c|c|}
\hline
\textbf{Scale}&\multicolumn{4}{c|}{\textbf{Number of positive examples $nop$}} \\
\cline{2-5} 
\textbf{$(L\times|\Sigma|)$} & 300 & 450 &600 &750 \\
\hline
$(7,3)$& $48.4/3$ & $67.5/3$ & $88.3/3$ & $113.6/3$\\
\hline
$(7,4)$& $85.3/3$ & $124.7/3$ &$152.1/3$ &$214.9/3$\\
\hline
$(8,3)$& $334.5/3$ & $452.2/3$ &$581.0/3$ &$794.7/3$\\
\hline
$(8,4)$& $370.4/2$ & $489.6/2$ &$834.5/3$ &$1041/3$\\
\hline
$(9,3)$& $465.2/2$ & $681/2$ &$852.8/3$ &$1093/3$\\
\hline
$(9,4)$& $603.7/1$ & $824.6/1$ &$1317/2$ &$1560/2$\\
\hline
\end{tabular}
\label{tab:input}
\end{center}
\end{table}

We also conducted experiments to evaluate the number of examples needed to synthesize a TRE exactly matching the target one. We selected the TREs in table \ref{tab:TREs_1} with $L\in[7,9]$, alphabet size $|\Sigma|\in[3,4]$ and chose range of example length $l\leq 11$. As shown in Table~\ref{tab:scal}, 300 examples are not enough to synthesize a TRE exactly matching the target TRE. We extended the number of the positive examples $nop=300, 450, 600, 750$, and did the same for the negative example set. For each group, we still conducted 3 trials. Table~\ref{tab:input} presents the average running time and the number of trials generating matching TRE. For example, ``48.3/3'' means that the average running time is 48.3 seconds and all 3 trials are successful. The results show that as the length of TRE and the size of the alphabet increase, the number of examples required for a correct synthesis result increases significantly.

\subsection{Case study} \label{sbsc:case_study}

The first case study is to continue the illustrative example given in Section~\ref{sc:motivating}. For the gray-box setting, we can build the model with parameters directly as we did in Section~\ref{sc:motivating}. Therefore, we aim to see if our synthesis approach helps to extract the parameters from sampled traces. For sampling traces, the parameters of the case are set as: minimum interarrival times $ p_1 = 8$, $ p_2 = 14$, execution times $C_1 = 4$, $ C_2 = 6$, and deadlines $D_1=8,$ $D_2=20$.


\begin{table}[!t]
\caption{The template $p$TREs in Scheduling case study.}
\label{tab:schedtemp}
\begin{center}
\begin{tabular}{|c|p{4.2cm}<{\centering}|p{1cm}<{\centering}|p{1.3cm}<{\centering}|}
\hline
\textbf{No.}& \textbf{Template $p$TRE}& \textbf{States}& \textbf{Parameters} \\
\hline
1  &\makecell{$\tr{\underline{a_2}_{\mathbb{I}_1}\cdot\underline{a_1}_{\mathbb{I}_2}}_{\mathbb{I}_3}\cdot\underline{b_1}$\\$\cdot\tr{\tr{\underline{a_2}\cdot\underline{b_2}}_{\mathbb{I}_4}}^*\cdot \underline{b_2}$} &\makecell{$q_0,q_2$, \\$q_3,q_5$}&\makecell{$p_1,p_2$, \\$D_2-D_1$} \\
\hline
2  &\makecell{$\tr{\underline{a_1}_{\mathbb{I}_1}\cdot\underline{b_1}}\vee\tr{\underline{a_2}_{\mathbb{I}_2}\cdot\underline{b_2}}$ \\ $\vee\tr{\tr{\underline{a_2}_{\mathbb{I}_3}\cdot\underline{a_1}}_{\mathbb{I}_4}\cdot\underline{b_2}\cdot\underline{b_1}}$} &\makecell{$q_0,q_1$, \\$q_2,q_4$}&\makecell{$p_1,p_2$, \\ $D_2-D_1$} \\
\hline
3  &\makecell{$\tr{\underline{a_1}_{\mathbb{I}_1}\cdot\underline{b_1}}\vee\tr{\underline{a_2}_{\mathbb{I}_2}\cdot\underline{b_2}}\vee\tr{\underline{a_1}_{\mathbb{I}_3}}$ \\ $\cdot\underline{a_2}_{\mathbb{I}_4}\cdot\tr{\tr{\underline{b_1}\cdot\underline{a_1}}_{\mathbb{I}_5}^*}_{\mathbb{I}_6}\cdot\underline{b_1}\cdot\underline{b_2}$} &\makecell{$q_0,q_1$, \\$q_2,q_3$}&\makecell{$p_1,p_2,$ \\ $D_2-D_1$} \\
\hline
\end{tabular}

\end{center}
\end{table}

\begin{table}[tb]
\caption{The Running Time (seconds) for Template $p$TREs.\\$\textbf{Above}$: Template 1. $\textbf{Middle}$: Template 2. $\textbf{Below}$: Template 3.}
\label{tab:schedtime}
\begin{center}
\begin{tabular}{|c|c|c|c|c|}
\hline
\textbf{Length of}&\multicolumn{4}{c|}{\textbf{Number of positive examples $nop$}} \\
\cline{2-5} 
\textbf{Examples $l$} & 300 & 450 &600 &750 \\
\hline
$\le11$&$42.4$  & $56.3$ & $73.8 $ & $95.5 $\\
\hline
$\le15$& $52.3$ & $75.2$ &$98.5$ & $130.6$\\
\hline
$\le19$& $59.6$ & $84.7$ &$113.1$ &$152.8$\\
\hline
\end{tabular}

\vspace{0.5cm}
\begin{tabular}{|c|c|c|c|c|}
\hline
\textbf{Length of}&\multicolumn{4}{c|}{\textbf{Number of positive examples $nop$}} \\
\cline{2-5} 
\textbf{Examples $l$} & 300 & 450 &600 &750 \\
\hline
$\le11$&$58.7$  & $75.5$ & $94.6 $ & $116.3 $\\
\hline
$\le15$& $67.9$ & $92.2$ &$114.5$ & $ 138.3$\\
\hline
$\le19$& $86.0$ & $118.7$ &$143.6$ &$177.1$\\
\hline
\end{tabular}

\vspace{0.5cm}
\begin{tabular}{|c|c|c|c|c|}
\hline
\textbf{Length of}&\multicolumn{4}{c|}{\textbf{Number of positive examples $nop$}} \\
\cline{2-5} 
\textbf{Examples $l$} & 300 & 450 &600 &750 \\
\hline
$\le11$&$86.8^a$  & $122.5$ & $157.4 $ & $183.6$\\
\hline
$\le15$& $117.3$ & $146.2$ &$174.7$ & $203.2$\\
\hline
$\le19$& $132.8$ & $169.0$ &$204.3$ &$241.5$\\
\hline
\end{tabular}
\end{center}
$^{\mathrm{a}}$ Not successfully generate all the target intervals.
\end{table}

We predefine three templates $p$TREs as listed in Table~\ref{tab:schedtemp}, each capturing a part of the automaton. We sample these templates $p$TREs with all time intervals to be $(0,30]$. For instance, the 4 target intervals in Template 1 are $\mathbb{I}_1=[p_2,30]$, $\mathbb{I}_2=(0,D_2-D_1]$, $\mathbb{I}_3=[p_1,30]$, $\mathbb{I}_4=[p_2,30]$. After sampling on the templates, we select the traces that can be accepted by the TA as positive examples, and the remaining as negative examples. Then we try to synthesize time intervals directly from these templates, skipping the enumeration of $p$TRE$_+$. We range the length of examples to $l\leq11,15,19$, the number of examples to $nop=300, 450, 600,750$. We conducted one trial for each template with each setting of $(l,nop)$. Table~\ref{tab:schedtime} presents the synthesis for each trial. In most cases (except for the case $l\leq 11$ and $nop=300$), our method successfully learned the parameters involved. For instance, it takes 42.4 seconds to get the time intervals $\mathbb{I}_1=[14,30]$, $\mathbb{I}_2=(0,12]$, $\mathbb{I}_3=[8,30]$, and $\mathbb{I}_4=[14,30]$ for Template 1 with $l\leq 11, nop=300$. So we can conclude from the synthesis results that $p_1=D_1=8$, $p_2=14$, $D_2-D_1=12$, and $D_2=20$. Then, according to EDF, we know that if we can guarantee the execution times $C_1, C_2$ satisfy $\frac{C_1}{8}+\frac{C_2}{14}\leq 1$, then the system is schedulable.

\begin{figure}
    \centering
    \resizebox{0.7\linewidth}{!}{
    \begin{tikzpicture}[->, >=stealth', shorten >=1pt, auto, node distance=5cm, semithick,scale=0.6, every node/.style={scale=0.8}, initial where=above]
    \node[initial,accepting,state] (0) {\large $q_0$};
    \node[state] (1) [right = 2cm of 0] {\large $q_1$};
    \node[state] (2) [below = 1cm of 1] {\large $q_2$};
    \node[state] (3) [below left = 1.3cm of 2] {\large $q_3$};
    \node[state] (4) [below =1cm of 0] {\large $q_4$};
    \path (0) edge node [above] {\large $a,\top,\{x\}$} (1)
    (1) edge  node [right] {\large $b,x\ge2,\{x\}$} (2)
    (2) edge node [above] {\large $d,x\ge12,\{x\}$} (4)
    (4) edge node [left] {\large $e,x\ge3,\{\}$} (0)
    (2) edge node [right] {\large $c,\top,\{x\}$} (3)
    (3) edge node [left] {\large $d,x\ge7,\{x\}$} (4);
    \end{tikzpicture}
    }
    \caption{Train TA}
    \label{fig:TrainTA}
\end{figure}
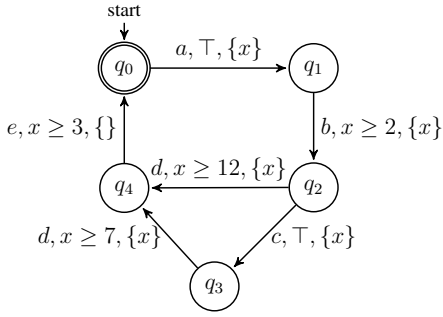

\begin{table}[!t]
\caption{The Running Time (seconds) of Synthesizing TRE in Train TA Experiment.}
\label{tab:TrainTA}
\begin{center}
\begin{tabular}{|c|c|c|c|c|}
\hline
\textbf{Length of}&\multicolumn{4}{c|}{\textbf{Number of positive examples $nop$}} \\
\cline{2-5} 
\textbf{Examples $l$} & 300 & 450 &600 &750 \\
\hline
$\le11$&$3347$  & $3756$ & $6882/\bigstar $ & TO\\
\hline
$\le15$& $1825$ & $3912$ &$4458$ & TO\\
\hline
$\le19$& $1772$ & $1946$ &$4751$ &$5239$\\
\hline
\end{tabular}
\end{center}
$\bigstar$ indicates the resulted TRE exactly matches the target TRE $\varphi_{\textit{Train}}$;\\ TO represents a timeout.
\end{table}

The second case study is about a train gate simplified from  \cite{10.1007/978-3-031-06773-0_26}, which is modeled by the TA as depicted in Fig. \ref{fig:TrainTA},  where $a,b,c,d,e$ stand for ``starts'', ``train approaches'', ``train halts'', ``train enters the gate'', and ``train leaves'', respectively. The corresponding TRE of this TA is $\varphi_{\textit{Train}}=\tr{\underline{a}\cdot\tr{\underline{b}}_{[2,\infty)}\cdot\tr{\tr{\underline{d}}_{[12,\infty)}\vee\tr{\underline{c}\cdot\tr{\underline{d}}_{[7,\infty)}}}\cdot\tr{\underline{e}}_{[3,\infty)}}^*$. It is more complex than the ones in previous subsections with length $L=12$ and alphabet size $|\Sigma|=5$. We set the range of examples length to $l\leq11,15,19$, the number of examples to $nop=300, 450, 600,750$, and the maximum delay time to 20 (substituting for $\infty$ as time upper bounds). We conducted one trial for each setting with a time limit of 7200 seconds. Table~\ref{tab:TrainTA} presents the running time for each trial. In most cases, our method terminates and returns the minimal TRE consistent with the randomly generated examples. What's more, a trial even returns a TRE exactly matching the target TRE. It shows the effectiveness and efficiency of our method in practice.

\section{Concluding Remarks} \label{sc:conclusion}
We addressed the TRE synthesis problem by theoretically establishing its decidability and proposing a practical framework aimed at finding a minimal TRE recognizing  the given set of real-time behaviors. The framework is based on the insight of first finding a $p$TRE that accepts all positive examples and then appropriately deciding its time intervals. We have implemented the synthesis method, along with three strategies for enumerating and pruning $p$TREs. The three strategies were compared experimentally and the framework's scalability was demonstrated through tests on uniformly sampled real-time behaviors of TAs, which are from TREs of various complexities. 

Note that, as we mentioned in Section~\ref{sc:preliminary}, although (generalized extended) timed regular expressions have the same expressive power as timed automata~\cite{AsarinCM02}, we only consider (extended) timed regular expressions in this work, excluding the renaming function ($\theta$) and the conjunction ($\wedge$). Therefore, our method 
thus may not apply to all real-time scheduling models, which are beyond extended timed regular expression. Which kinds of timed automata have the same expressive power as (extended) timed regular expressions remains  unclear.

In future work, we plan to explore other enumeration and pruning strategies for our framework. The challenge lies in striking a balance between reducing the effort required to check $p$TREs and expanding the scope of pruning effectively. Additionally, we intend to identify the sufficient conditions for positive (and negative) examples required by our synthesis framework.  
Furthermore, we are to provide a PAC bound regarding the number of samples needed to achieve sufficient conditions of examples incorporating the uniform sampling technique. Which kinds of scheduling models can be expressed by (extended) timed regular expressions is also an interesting research topic. 

\section*{Acknowledgment}
We thank the anonymous reviewers for their valuable comments, which significantly improved the quality of this paper. Ziran Wang, Jie An, and Naijun Zhan are supported by the National Key R\&D Program of China under grant No.~2022YFA1005101, the Natural Science Foundation of China (NSFC) under grants No.~W2511064, No.~62192732 and No.~62032024, and the ISCAS Basic Research Grant No.~ISCAS-JCZD202406. Miaomiao Zhang is supported by the NSFC Grant No.~62472316. Zhenya Zhang is supported by the JSPS Grant No.~JP25K21179.


\bibliographystyle{IEEEtran}
\bibliography{reference}

\newpage
\appendices
\section{Proofs of Lemmas and Theorems}
\subsection{Proof of Lemma \ref{lemma:sEL}}
\begin{proof}
Suppose $\omega_1 = (\sigma_1^1,t_1^1)(\sigma_2^1,t_2^1) \cdots (\sigma_n^1,t_n^1)$, $\omega_2=(\sigma_1^2,t_1^2)(\sigma_2^2,t_2^2)\cdots(\sigma_m^2,t_m^2)$, and $\mathcal{SE}(\omega_1)=\mathcal{SE}(\omega_2) $. 

Let us assume there is a TRE $\varphi$ distinguishing them such that $\omega_1\in \semantic{\varphi} \wedge \omega_2\not\in \semantic{\varphi}$. 

By Definition \ref{def:sel}, $\omega_1$ and $\omega_2$ have $(\sigma_1^1,\sigma_2^1,\cdots,\sigma_n^1)=(\sigma_1^2,\sigma_2^2,\cdots,\sigma_m^2)$. And for $\forall 1\le j\le k\le n$, either $\sum\nolimits_{i=j}^{k}{t_i^1},\\\sum\nolimits_{i=j}^{k}{t_i^2}\in (d,d+1)$ or $\sum\nolimits_{i=j}^{k}{t_i^1}=\sum\nolimits_{i=j}^{k}{t_i^2} = d$, and $d \in \mathbb{N}$. 

Consider a time constraint $\sum\nolimits_{i=j}^{k}{t_i} \in I =$$\tr{u,v}$ of $\semantic{\varphi}$, where $\langle\in \{[,(\}$ and $\rangle\in\{],)\}$, $u,v \in \mathbb{N}, 1\le j\le k\le n$. If $\sum\nolimits_{i=j}^{k}{t_i^1}=\sum\nolimits_{i=j}^{k}{t_i^2} = d$, then obviously $\sum\nolimits_{i=j}^{k}{t_i^1} \in I \iff \sum\nolimits_{i=j}^{k}{t_i^2} \in I$. And in the case $\sum\nolimits_{i=j}^{k}{t_i^1},\sum\nolimits_{i=j}^{k}{t_i^2}\in (d,d+1)$ and $\sum\nolimits_{i=j}^{k}{t_i^1} \in I$, we can see that if $u > d$ or $ v <d+1$, then $(d,d+1)\cap I = \emptyset$, which is a contradiction. It must be the case $ u\leq d<d+1\leq v$, and  $(d,d+1) \subseteq I$. Hence, $\sum\nolimits_{i=j}^{k}{t_i^1} \in I \Rightarrow \sum\nolimits_{i=j}^{k}{t_i^2} \in (d,d+1) \subseteq I$. $\sum\nolimits_{i=j}^{k}{t_i^2} \in I \Rightarrow \sum\nolimits_{i=j}^{k}{t_i^1} \in (d,d+1) \subseteq I$ is likewise. 

Therefore, $\omega_2$ satisfies any action or time constraints of $\phi$ that $\omega_1$ satisfies, which is a contradiction. 
\end{proof}

\subsection{Proof of Lemma \ref{lemma:num_sTRE}}
\begin{proof}
Given an $n$-length timed word $\omega_0$, we need to prove that the semantic of a sTRE $\varphi_{\omega_0}\in\Phi(\omega_0)$ can be expressed in the form $\semantic{\varphi}=\{\omega|\mu(\omega)=\mu(\omega_0)\}\cap\{\omega|\lambda(\omega)\models\theta_1\wedge\theta_2\wedge\cdots\wedge\theta_l,\theta_{1,2,\cdots,l}\in\Theta(\omega_0)\}$, where $\omega$ is a timed word.

This is clearly valid at the length of $\omega_0$ to be $1$. 
Suppose the lemma is valid at the length of $\omega_0$ $\leq k$. If the length of $\omega_0$ is $k_0=k+1$, suppose $\omega_0=(\sigma_1^0,t_1^0,1)(\sigma_2^0,t_2^0)\cdots(\sigma_k^0,t_k^0)(\sigma_{k+1}^0,t_{k+1}^0)$. 

If $\varphi_{\omega_0}\in\Phi(\omega_0)$ is of the form $\varphi_1\cdot\varphi_2$, where the lengths of $\varphi_1$ and $\varphi_2$ are $k_1,k_2\leq k$, $k_2=k+1-k_1$, then by Definition \ref{def:stre}, $\varphi_1$ and $\varphi_2$ contain no operation other than $\cdot$ and $\tr{}_{I}$, and all time restrictions $I$ in them are in the form of $(d,d+1)$ or $[d,d],~d\in\mathbb{N}$. So $\varphi_1$ and $\varphi_2$ are both sTREs. 
We denote $\omega_1=(\sigma_1^0,t_1^0)(\sigma_2^0,t_2^0)\cdots(\sigma_{k_1}^0,t_{k_1}^0)$ and $\omega_2=(\sigma_{k_1+1}^0,t_{k_1+1}^0,1)(\sigma_{k_1+2}^0,t_{k_1+2}^0)\cdots(\sigma_k^0,t_k^0)$\newline$(\sigma_{k+1}^0,t_{k+1}^0)$. Since an sTRE accepts only timed words of the same length as itself, we have $\omega_1\in\semantic{\varphi_1}$ and $\omega_2\in\semantic{\varphi_2}$. Then $\semantic{\varphi_1}=\{\omega|\mu(\omega)=\mu(\omega_1)\}\cap\{\omega|\lambda(\omega)\models\theta_1\wedge\theta_2\wedge\cdots\wedge\theta_{l_1},\theta_{1,2,\cdots,_{l_1}}\in\Theta(\omega_1)\}$ and $\semantic{\varphi_2}=\{\omega|\mu(\omega)=\mu(\omega_2)\}\cap\{\omega|\lambda(\omega)\models\theta_1\wedge\theta_2\wedge\cdots\wedge\theta_{l_2},\theta_{1,2,\cdots,_{l_2}}\in\Theta(\omega_2)\}$. And by Definition \ref{def:tre}, $\semantic{\varphi_0}=\{\omega|\mu(\omega)=\mu(\omega_1\cdot\omega_2)\}\cap\{\omega|\lambda(\omega)\models\theta_1\wedge\theta_2\wedge\cdots\wedge\theta_{l_1}\wedge\theta'_1\wedge\theta'_2\wedge\cdots\wedge\theta'_{l_2},\theta_{1,2,\cdots,l_1}\in\Theta(\omega_1),\theta'_{1,2,\cdots,l_2}\in\Theta'(\omega_2)\}$, where $\theta'_i$ and $\Theta'(\omega_2)$ are derived by renaming the time constrains $\Theta(\omega_2)$ with $1\leq u\leq v\leq k_2$ to be ${k_1}+1\leq u\leq v\leq k+1$ in $\sum\nolimits_{i=u}^{v}{t_i}\in (d,d+1)$ and $\sum\nolimits_{i=u}^{v}{t_i}=d$. In this way $\omega_0=\omega_1\cdot\omega_2$, $\Theta(\omega_1),\Theta'(\omega_2)\subset\Theta(\omega_0)$, so $\semantic{\varphi_0}$ is expressed in the form we want.

If $\varphi_{\omega_0}\in\Phi(\omega_0)$ is of the form $\tr{\varphi'}_I$, where $\varphi'$ is another $(k+1)$-length TRE. Then $\varphi'$ is a sTRE and $\omega_0 \in \semantic{\varphi}\subset\semantic{\varphi'}$. So $\varphi'$ is also a sTRE of $\omega_0$, and $\varphi'$ is of the form $\varphi_1\cdot\varphi_2$. Then we have $\semantic{\varphi'}$ expressible in the form we want. So $\semantic{\varphi} = \semantic{\varphi'}\cap\{\sum\nolimits_{i=1}^{k+1}t_i\in I\}$ is also the form we want.

The number of time constraints in $\Theta(\omega_0)$ of $\mathcal{SE}(\omega_0)$ equals the number of $\sum\nolimits_{i=j}^{k}{t_i}, 1\leq j \leq k \leq n$, which is $(n^2+n)/2$, so the number of different  $\varphi_{\omega_0}$ is at most $2^{(n^2+n)/2}$. 
\end{proof}

\subsection{Proof of Lemma \ref{lemma:relation_sEL_sTRE}}
\begin{proof}
Given a timed word $\omega_1 = (\sigma_1^1,t_1^1)(\sigma_2^1,t_2^1)\cdots(\sigma_n^1,t_n^1)$, We have $\Lambda(\omega_1) = \bigwedge_{\theta\in\Theta(\omega_1)}\theta$. For any time constraint $\theta\in\Theta(\omega_1)$,  
$\lambda(\omega_1) \models \theta$. Suppose $\theta$ is of the form $d < \sum\nolimits_{i=j}^{k}{t_i}< d+1$ (or $\sum\nolimits_{i=j}^{k}{t_i} = d$), $1\leq j\leq k\leq n$ and $d \in \mathbb{N}$, then $\{\mu(\omega)=\mu(\omega_1)\}\cap\{\lambda(\omega) \models \theta\}$ is the semantic of TRE $\varphi = \underline{\sigma_1^1}\cdot\underline{\sigma_2^1}\cdot\cdots\cdot\underline{\sigma_{i-1}^1}\cdot\tr{\underline{\sigma_i^1}\cdot\underline{\sigma_{i+1}^1}\cdot\cdots\cdot\underline{\sigma_{j-1}^1}\cdot\underline{\sigma_j^1}}_{(d,d+1)(\text{or }[d,d])}\cdot\underline{\sigma_{j+1}^1}\cdot\cdots\cdot\underline{\sigma_n^1}$. $\varphi$ is a sTRE of $\omega_1$ by Definition \ref{def:stre}. So $\mathcal{SE}(\omega_1)= \{\mu(\omega)=\mu(\omega_1)\}\cap\bigcap_{\theta\in\Theta(\omega_1)}\{\lambda(\omega) \models 
\theta\}\supseteq\bigcap_{\varphi_{\omega_1}\in\Phi(\omega_1)}{\semantic{\varphi_{\omega_1}}}$.

By Lemma \ref{lemma:sEL}, $\forall \omega_2 \in \mathcal{SE}(\omega_1),\ \forall \varphi_{\omega_1}\in\Phi(\omega_1), \omega_2\in\semantic{\varphi_{\omega_1}} $. So $\mathcal{SE}(\omega_1)\subseteq\bigcap_{\varphi_{\omega_1}\in\Phi(\omega_1)}{\semantic{\varphi_{\omega_1}}}$. 
\end{proof}

\subsection{Proof of Lemma \ref{lemma:fals_dcd}}
\begin{proof}
 From lemma \ref{lemma:num_sTRE}, it takes at most $2^{(n^2+n)/2}$ TRE acceptance queries to check if $\exists \varphi_\omega \in \Phi(\omega), \omega'\in{\Omega}$, $\omega' \in \semantic{\varphi_\omega}$. So it takes at most $2^{(n^2+n)/2}\cdot m$ TRE acceptance queries to check all $m$ timed words in $\Omega$.  
\end{proof}

\subsection{Proof of Lemma \ref{lemma:stre_fals}}
\begin{proof}
By parsing how $\omega$ is accepted by $\phi$ we yield $\lambda \models \Lambda^{\phi}({\omega})$, where $\Lambda^{\phi}({\omega})$ is the set of inequalities (and equalities) about $\lambda(\omega)$ in current parsing situation.
Then there exists a sTRE $\varphi$ of $\omega$, whose time constraints $\Lambda^{\varphi}({\omega})$ is of the same form of $\Lambda^{\phi}({\omega})$, with only differences on the numbers in the inequalities (and equalities) that come from the interval bounds in $\phi$ and $\varphi$. Suppose one of the inequalities in $\Lambda^{\phi}({\omega})$, the corresponding inequality in $\Lambda^{\varphi}({\omega})$ must be the same or more strict by the definition of sTRE. So $\semantic{\varphi}\subseteq\semantic{\phi}$.
Then by the definition of obscuration, $\exists \omega_{-}\in{\Omega_{-}}, \omega_{-}\in\semantic{\varphi}\subseteq\semantic{\phi}$.
\end{proof}

\subsection{Proof of Theorem \ref{thm:decidability}}
\begin{proof}
If $\forall \omega_{+}\in{\Omega_{+}}$ is not obscured by $\Omega$, then for each $\omega_+$, there exists a sTRE $\varphi\in\Phi(\omega_+)$ such that $\forall\omega_{-}\in{\Omega_{-}},\omega_-\not\in\semantic{\varphi}$, which we denote as $\varphi_t(\omega_+)$ here. $\varphi=\bigvee_{\omega_+\in \Omega_+}\varphi_t(\omega_+)$ is a solution of Problem \ref{prob:syn_p_n}. 

If there exists $\omega_{+}\in{\Omega_{+}}$ that is obscured by $\Omega_-$, and suppose $\varphi'$ is a solution of Problem \ref{prob:syn_p_n}, then $\omega_{+}\in\semantic{\varphi'}$. By Lemma~\ref{lemma:stre_fals}, there exists $\omega_{-}\in{\Omega_{-}}, \omega_{-}\in\semantic{\varphi'} $, which is a contradiction. 
\end{proof}

\subsection{Proof of Lemma \ref{lemma:ptre_finite}}
\begin{proof}
Starting with $k=1$, the number of $p$TRE is obviously finite. 

If the number of $p$TRE is finite at $k\leq n$, then at $k=n+1$, any $(n+1)$-length $p$TRE can be generated by applying $
\phi := \tr{\phi^*}_{\mathbb{I}} $ on a $n$-length $p$TRE or by applying $\phi := \tr{\phi\cdot\phi}_{\mathbb{I}} \;\vert\; \tr{\phi\vee\phi}_{\mathbb{I}}$ on a $(n-1)$-length $p$TRE. The number of $\phi$ in a $n$-length (or $(n-1)$-length) $p$TRE is finite, and applying $\phi := \tr{\phi^*}_{\mathbb{I}} $ (or $\phi := \tr{\phi\cdot\phi}_{\mathbb{I}} \;\vert\; \tr{\phi\vee\phi}_{\mathbb{I}}$) on a particular $\phi$ yields 1 (or 2) particular $p$TRE. So the number of $(n+1)$-length $p$TRE finite. By induction, $\forall k \in \mathbb{N}$, the number of $k$-length $p$TRE is finite.
\end{proof}

\subsection{Proof of Lemma \ref{lemma:no_epsilon}}
\begin{proof}
Since $\Omega$ is nonempty,  $\varepsilon$ itself cannot be the minimal TRE. We then prove the lemma by showing that a TRE $\varphi\neq \varepsilon$ with $\varepsilon$ in it has the same semantics as a shorter TRE, so $\varphi$ cannot be the minimal TRE w.r.t $\Omega$.
\begin{itemize}
    \item If there is an $\varepsilon$ in $\tr{\varepsilon^*}_{\mathbb{I}} $, then by replacing this $\tr{\varepsilon^*}_{\mathbb{I}}  $ structure with $\varepsilon$ we yield a shorter $p$TRE $\varphi'$. By Definition \ref{def:tre}, $\semantic{\tr{\varepsilon^*}_{\mathbb{I}}}=\semantic{\varepsilon}=\{\varepsilon\}$, so $\semantic{\varphi}= \semantic{\varphi'}$.
    \item If there is an $\varepsilon$ in $\tr{\varepsilon\cdot\varphi_1}_{\mathbb{I}}$, where $\varphi_1$ is a $p$TRE. then by replacing this structure with $\tr{\varphi_1}_{\mathbb{I}}$ we yield a shorter $p$TRE $\varphi'$. By Definition \ref{def:tre}, $\semantic{\varepsilon\cdot\varphi_1}_{\mathbb{I}}=\semantic{\tr{\varphi_1}_{\mathbb{I}}}$, so $\semantic{\varphi}= \semantic{\varphi'}$.
    \item If there is an $\varepsilon$ in $\tr{\varepsilon\vee\varphi_1}_{\mathbb{I}}$, where $\varphi_1$ is a $p$TRE. then by replacing this structure with $\tr{\varphi_1}_{\mathbb{I}}$ we yield a shorter $p$TRE $\varphi'$. By Definition \ref{def:tre}, $\semantic{\varepsilon\vee\varphi_1}_{\mathbb{I}}=\semantic{\tr{\varphi_1}_{\mathbb{I}}}$, so $\semantic{\varphi}= \semantic{\varphi'}$.
\end{itemize} 
\end{proof}

\subsection{Proof of Lemma \ref{lemma:trivial_gen}}
\begin{proof}
We mark the initial $\hole$ with serial number $0$, and the $i$th $\hole$ introduced in $p$TRE generation with $i$ (for $\hole\rightarrow\tr{\hole\cdot\hole}_{\mathbb{I}}|\tr{\hole\vee\hole}_{\mathbb{I}}$, the serial number of the right $\hole$ is larger). A sequence $\overline{s}=\overline{s_1 s_2\cdots s_n}$ is composed of generating steps $s_1,s_2,\cdots,s_n$, where $\forall1\leq j \leq n,s_j=(i_j, g_j), i_j\in \mathbb{N}$ denoting the serial number of $\hole$ applied generating rule $g_j\in \hole\rightarrow\tr{\underline{\sigma}}_{\mathbb{I}}|\tr{\hole^*}_{\mathbb{I}}|\tr{\hole\cdot\hole}_{\mathbb{I}}|\tr{\hole\vee\hole}_{\mathbb{I}}$. We can see a sequence is feasible iff the $i_j$th $\hole$ is already introduced and is not yet consumed when $s_j=(i_j,g_j)$ is taken. If a sequence is feasible, it actually generates a $p$TRE, and every $p$TRE has at least one feasible sequence to generate it. We say $\overline{s_1}=\overline{s_2} $ iff the two sequences are both feasible and generate the same $p$TRE.

We need to prove the commutative law of $\hole\rightarrow\tr{\underline{\sigma}}_{\mathbb{I}}$ in feasible sequences, i.e. $\overline{s_j,s_{j+1}}=\overline{s_{j+1},s_j}$ if $s_j=(i_j,g_j), g_j \in \hole\rightarrow\tr{\underline{\sigma}}_{\mathbb{I}}$. $s_j=(i_j,g_j)$ is available at the $j$th step iff the $\hole$ $i_j$ is introduced and hasn't been consumed by another generating step. And since $\hole$ $i_j$ is consumed by $s_j$, it cannot be consumed by $s_{j+1}$, i.e. $i_j\neq i_{j+1}$. Besides, $s_j$ doesn't introduce new $\hole$, so the $\hole$ consumed by $s_{j+1}$ is introduced before $s_j$. So we can see the sequence is still feasible after switching $s_j$ and $s_{j+1}$. And it can be verified that $\hole$ $i_j$ and $i_{j+1}$ are applied the same generating rules regardless of switching or not.

Suppose that a $k$-length $p$TRE is generated by a feasible sequence $\overline{s}=\overline{s_1 s_2\cdots s_n}$. We rearrange the sequence to be $\overline{s'}=\overline{s'_1 s'_2\cdots s'_m s'_{m+1} \cdots s'_n}$, where $s'_{1,2,\cdots,m}$ are the steps applying $\hole\rightarrow\tr{\hole^*}_{\mathbb{I}}|\tr{\hole\cdot\hole}_{\mathbb{I}}|\tr{\hole\vee\hole}_{\mathbb{I}}$ and keeps their relative order in $\overline{s}$, and $\overline{s'_{m+1,m+2,\cdots,n}}$ are the steps applying $\hole\rightarrow\tr{\underline{\sigma}}_{\mathbb{I}}$ in $\overline{s}$. With the commutative law above, we have $\overline{s}=\overline{s'}$. By Algorithm \ref{algorithm: trivial}, the result of sequence $\overline{s'_1 s'_2\cdots s'_m}$ is contained in $\mathit{S_k}$, and then result after $\overline{s'_{m+1,m+2,\cdots,n}}$ is contained in $\mathit{S'_k}$.
Therefore, all $p$TREs are to be generated.
\end{proof}

\subsection{Proof of Theorem \ref{thm:recursive_gen_len}}
\begin{proof}
 Assuming in the generation of a $p$TRE $\phi$, ${\hole\rightarrow \tr{\hole\cdot\hole}_{\mathbb{I}}}$ and ${\hole\rightarrow \tr{\hole\vee\hole}_{\mathbb{I}}}$ are used for $a$ times in total, $\hole\rightarrow \tr{\hole^*}_{\mathbb{I}}$ for $b$ times, and $\hole\rightarrow \tr{\sigma}_{\mathbb{I}}$ for $c$ times. Each time ${\hole\rightarrow \tr{\hole\cdot\hole}_{\mathbb{I}}}$ or ${\hole\rightarrow \tr{\hole\vee\hole}_{\mathbb{I}}}$ is used, the length of the $p$TRE after this substitution is 2 more than before. Likewise, the length increases by 1 after each use of $\hole\rightarrow \tr{\hole^*}_{\mathbb{I}}$. $\hole\rightarrow \tr{\underline{\sigma}}_{\mathbb{I}}$ do nothing with the length. So the length of $\phi$ is $k=2a+b+1$. In the other hand, considering the number of $\hole$s, each time ${\hole\rightarrow \tr{\hole\cdot\hole}_{\mathbb{I}}}$ or ${\hole\rightarrow \tr{\hole\vee\hole}_{\mathbb{I}}}$ is used, the number of $\hole$s increases by one, while $\hole\rightarrow \tr{\hole^*}_{\mathbb{I}}$ doesn't increase the number and $\hole\rightarrow \tr{\underline{\sigma}}_{\mathbb{I}}$ eliminate 1 placeholder. The number of $\hole$s starts with 1 and ends with 0, so $a+1-c=0$. The number of steps taken is then $a+b+c=2a+b+1=k$. So the closed $p$TREs, and hence the $p$TRE$_+$ generated after $k$ recursive steps are of length $k$. Reversely, a $p$TRE$_+$ of length $k$ can only be generated at exactly the $k$-th step. So all $k$-length $p$TRE$_+$ are generated at the $k$-th step in the recursive generation.
 \end{proof}

\subsection{Proof of Theorem \ref{thm:ptre_formula_sat}}
\begin{proof}
If $\Phi_{\Omega}$ is satisfiable, then one of its models is an instance of $\phi$, e.g., the interval parameters of a TRE $\varphi\in[\phi]$. In this model, for each $\omega_+\in\Omega_+$, at least one accepting path formula of $\omega_+$ w.r.t $\phi$ is satisfied. So $\omega_+\in\semantic{\varphi}$. For each $\omega_-\in\Omega_-$, if $\omega_-\not\in\semantic{\phi}$, then $\omega_-\not\in\semantic{\varphi}$. If $\omega_-\in\semantic{\phi}$, then $\omega_-$ is parsed and the accepting paths are encoded. But by definition of $\Phi_{\Omega}$, none of these path formulas is satisfied. So $\omega_-\not\in\semantic{\varphi}$.

If there exists a TRE $\varphi\in[\phi]$ consistent with $\Omega=(\Omega_+,\Omega_-)$, then for each $\omega_+\in\Omega_+$, $\omega_+\in\semantic{\varphi}$, meaning there is at least an accepting path formula of $\omega_+$ w.r.t $\phi$ satisfied by $\varphi$'s interval parameters. And for each $\omega_-\in\Omega_-$, either $\omega_-\not\in\semantic{\phi}$, then $\omega_-$ doesn't appear in $\Phi_{\Omega}$. Or $\omega_-\in\semantic{\phi}$ but $\omega_-\not\in\semantic{\varphi}$, meaning none of the accepting path formulas of $\omega_-$ w.r.t $\phi$ is satisfied by $\varphi$'s interval parameters. Then, by definition of $\Phi_{\Omega}$, we have $\varphi$'s interval parameters to be a model of it.
\end{proof}

\end{document}